\documentclass[nofootinbib,notitlepage,aps,pra,twocolumn,groupaddress,superscriptaddress,10pt]{revtex4-1}

\usepackage[english]{babel}

\usepackage{graphicx}
\usepackage[colorlinks=true, allcolors=blue]{hyperref}

\usepackage{stmaryrd}
\usepackage{amssymb,amsmath,amsthm,amsfonts,amsbsy}
\usepackage{bm,bibunits,color,chngcntr,epsfig,epstopdf,graphicx,dsfont}
\usepackage{hyperref,lipsum,,makecell,mathrsfs,rotating,setspace}
\usepackage[english]{babel}
\usepackage[normalem]{ulem}
\usepackage{array}
\usepackage{makecell} 

\newcommand{\be}{\begin{equation}}
\newcommand{\ee}{\end{equation}}
\newcommand{\bea}{\begin{eqnarray}}
\newcommand{\eea}{\end{eqnarray}}
\newcommand{\ket}{\rangle}
\newcommand{\bra}{\langle}

\newcommand{\I}{\mathds{1}}

\newcommand{\ba}{\begin{align}}
\newcommand{\ea}{\end{align}}
\newcommand{\Tr}{\text{tr}}

\def\C#1{\mathcal #1}

\begin{document}

\title{Towards transistor-based quantum computing}
\author{Yuan-Dong Liu}
\thanks{These authors contributed equally to this work.}
\affiliation{Institute of Theoretical Physics, Chinese Academy of Sciences, Beijing 100190, China \\
School of Physical Sciences, University of Chinese Academy of Sciences, Beijing 100049, China}

\author{Xiang Xu}
\thanks{These authors contributed equally to this work.}
\affiliation{Institute of Theoretical Physics, Chinese Academy of Sciences, Beijing 100190, China \\
School of Physical Sciences, University of Chinese Academy of Sciences, Beijing 100049, China}

\author{Qing-Rui Wang}
\affiliation{Yau Mathematical Sciences Center, Tsinghua University, Haidian, Beijing 100084, China}

\author{Dong-Sheng Wang}
\thanks{Corresponding author}
\email{wds@itp.ac.cn}
\affiliation{Institute of Theoretical Physics, Chinese Academy of Sciences, Beijing 100190, China \\
School of Physical Sciences, University of Chinese Academy of Sciences, Beijing 100049, China}

\newtheorem{theorem}{Theorem}
\newtheorem{prop}[theorem]{Proposition}
\newtheorem{corollary}[theorem]{Corollary}
\newtheorem{open problem}[theorem]{Open Problem}
\newtheorem{conjecture}[theorem]{Conjecture}
\newtheorem{definition}{Definition}
\newtheorem{remark}{Remark}
\newtheorem{example}{Example}
\newtheorem{task}{Task}
\newtheorem{protocol}{Protocol}

\begin{abstract}
In this work, we propose and study in depth a universal quantum computing architecture
based on a quantum construction of transistors. 
Our teleportation-based quantum transistors, called ``telesistors'',
are ground states of systems with symmetry-protected topological order,
hence suppress certain noises and provide high-fidelity Clifford gates  
without the need for active error correction. 
This physical protection, quantified by the string order parameters, 
serves as a low-overhead foundation upon which 
conventional fault-tolerant encoding (e.g., with stabilizer codes) 
can be built to achieve universal quantum computation.
This architecture shows rich connections with current known architectures, 
and some desirable merits especially compared with the qubit-based circuits 
regarding modularity, integration, and program storage.
Our study shows that it is plausible to realize it
with current technology in the near future.
\end{abstract}
\date{\today}

\maketitle

\begin{spacing}{1.2}

\section{Introduction}

Modern quantum computing is based on qubits~\cite{NC00}. 
A quantum circuit or algorithm often 
starts from an initialization of qubits, 
and then gate operations on them, followed by final measurements.
This is a manifest of the so-called circuit model, 
as the analog of classical Boolean circuit.
However, this abstract model does not specify 
the spacetime features of both qubits and gates, 
e.g., how they are \textit{stored} in hardware,
or how the quantum data-flow and control-flow interacts. 
This offers diverse flexibility for developing 
physical quantum computing platforms~\cite{LJL+10},
as well as distinct universal quantum computing 
models~\cite{RB01,BK05,CGW13,CMP18,AL18,Arr19,YRC20,W24rev}.

Classical computers contain both external and internal memory units,
with the latter relying on transistors~\cite{HH13}.
For quantum computing, it is largely unknown how to construct 
the quantum analog of transistors and how to use them for 
universal quantum computing. 
To solve this, one must understand how to store quantum operations,
instead of states, in hardware. 
This has been advanced from the recent study of 
quantum von Neumann architecture (QvN)~\cite{W22_qvn},
which relies on the channel-state duality~\cite{Cho75} for gate storage
and measurement for its execution. 

In this work, we develop a quantum transistor-based architecture for 
universal quantum computing by operating on quantum transistors,
instead of qubits, extending our recent post~\cite{W26}. 
Different from the qubit-based architecture,
in our scheme both qubits and gates can be stored in quantum medium.
Table~\ref{tab:comp} compares the two architectures in brief,
and we can see that our transistor-based architecture integrates 
more quantum elements, hence offering novel opportunity 
for more broader development of quantum technology.

The mechanism to execute gates stored in quantum transistors
is gate teleportation~\cite{BBC+93,GC99,ZLC00}, 
which is also the case in the model of 
measurement-based quantum computing (MBQC)~\cite{RB01}. 
Therefore, our architecture combines ideas from MBQC and QvN,
as well as the circuit model,
with gate execution by measurement-induced gate teleportation
and gate storage via Choi states from channel-state duality.

This mirrors the traditional design of 
classical electronic circuits built from transistors~\cite{HH13}, 
where gates are physical components and bits are generated dynamically~\cite{note_bit}. 
Photonic quantum computing follows a somewhat similar approach, 
though its scalability is challenged by the probabilistic nature of entangling gates, 
which significantly increases overhead~\cite{KLM01}.
The idea of quantum transistors has previously been explored using
cluster states, but instead of using measurements,
it requires adiabatic control to make it reversible~\cite{BFC13,WB15}.
Another idea of quantum state transfer~\cite{AK25} has been explored,
which yet does not aim for universal quantum computing. 
In our scheme, the measurement-based gates must be reset for the next usage. 
This necessitates controllability over the gate Hamiltonian,
a requirement achievable with current technology. 
This ``one-time'' use is a common feature in quantum computing as 
the final measurement for readout irreversibly collapses quantum state.

For MBQC, it relies on a highly entangled resource state as a ``substrate,''
typically a cluster state, 
on which computation proceeds via adaptive single-qubit projective measurements.
Recently, a series of study~\cite{WSR17,RWP17,SWP+17,ROW+19,SNB+19,DAM20,Wei18,RYA23} 
revealed the close relation between MBQC
and resource states with symmetry-protected topological (SPT) order~\cite{GW09},
based on earlier effects of relating MBQC to 
ground states of quantum phases of matter~\cite{BM08,DB09,BBM10,Miy10,RMB13}. 
This approach generalizes the resource from a specific cluster state to an entire SPT phase, 
exploiting its universal properties to encode qubits and implement gates. 
Despite the protection of the edge modes, hence logical qubits, by symmetry, 
the proposed gates are not protected and their implementations are sophisticated.

By fault-tolerance consideration, 
we consider the discrete universal gate set of Hadamard gate $H$, 
phase gate $S$, $T$ gate, and $CZ$ gate.
A gate is stored as the ground state of a Hamiltonian $H(\lambda)$ 
for a system with edge modes,
and it is enabled by measurement. 
Each gate, as a system with edges, has a ``transistor-like'' structure:
the two edges are for input/output, and the bulk is for control. 
Before execution, a gate is in the form of Choi state, 
which is a program state central to QvN and protected by its SPT order.
Different from conventional MBQC, 
our architecture is \emph{modular},
and we only employ a semi-global measurement in a fixed SPT ``wire basis'' for each gate.
This allows straightforward correction of Pauli byproducts 
and eliminates the need for feed-forward of measurement outcomes.

We show that the transistors can be interconnected to form circuits.
To interconnect gates, 
additional qubits carried by quantum dots (QD) or other systems 
serving as junctions are used.    
The QD are designed to couple with the edge modes and couple among themselves. 
This brings controllability to the gate operations 
and modularity to the architecture.
It is also necessary for the $T$ gate, 
which is realized via the magic-state injection scheme~\cite{ZLC00,BK05}.
We also show that controllable coupling among QDs 
enables programmability of gate sequences.

The appearance of SPT order is not a coincidence. 
A recent study shows the intrinsic relation between teleportation and SPT order~\cite{HSF24}.
From the point of view of coding, 
our scheme establishes a new usage of SPT codes in quantum computing.
The usual MBQC scheme uses measurement-induced gates,
while the covariant codes~\cite{WZO+20,WWC+22,ZLJ20,YMR+22} uses symmetry-based unitary transversal gates.
Despite the difference,
a common feature is that the code distance of SPT code is small
due to the short-range or finite amount of entanglement, 
hence not enough to fully guarantee fault-tolerance.
This can be enhanced by concatenation, and we show that 
the transistors can be wired together to form coding structures.

\begin{table}[t!]
    \centering
    \footnotesize
    \renewcommand{\cellalign}{cc}  
    \begin{tabular}{|c|c|c|}
        \hline
        & \textbf{Qubit-based} & \textbf{Transistor-based} \\
        \hline
        \makecell{Circuit} & \makecell{combinational} & \makecell{combinational/\\ sequential} \\
        \hline
        \makecell{Computing\\ model} & \makecell{circuit model} & \makecell{QvN} \\
        \hline
        \makecell{Storage} & \makecell{quantum states} & \makecell{quantum states\\ and gates} \\
        \hline
        \makecell{Instruction} & \makecell{data-flow\\ driven} & \makecell{control-flow\\ driven} \\
        \hline
        \makecell{Program} & \makecell{linear} & \makecell{linear/\\ iterative} \\
        \hline
        \makecell{Classical\\ analog} & \makecell{magnetic-core\\ memory} & \makecell{CPU} \\
        \hline
    \end{tabular}
    \caption{A brief comparison between 
    the qubit-based and transistor-based quantum computing architecture.
    It emphasizes on the types of circuits, 
    relation with computing models, storage, instruction, storage of program, etc.
    Their distinction may not be a dichotomy, however,
    and they can also be combined together. 
    The construction of quantum sequential circuits is from our recent post~\cite{W26}.
    For classical computer system, please refer to Ref.~\cite{HH13}.
    }
    \label{tab:comp}
\end{table}

To demonstrate the SPT nature of the gates, 
we use matrix-product state (MPS) representation of states~\cite{PVW+07} in a cluster phase.
Improving upon prior results, 
we show that the gates $H$, $S$, and $CZ$ are protected by symmetry,
e.g., the $Z_2\times Z_2$ symmetry for 1D qubit cluster phase~\cite{ESB+12}.
Each gate requires a unique measurement pattern 
and also system size, 
a requirement normally absent in classical digital devices.
The $T$ gate is partially protected,
and its fidelity is affected by the QD-edge coupling.
For the $CZ$ gate, quasi-2D cluster phase on the square or honeycomb lattice
can be employed.

Our model admits several natural extensions. 
Beyond cluster phases, other SPT phases like the 1D Haldane phase 
and 2D valence-bond solids can be harnessed~\cite{AKLT87}. 
Promisingly, these phases are physically realizable in systems 
such as metal-organic magnetic chains~\cite{GPB+22,PEC+24,JCA+25},
quantum simulators~\cite{ERA25} 
or via Floquet engineering~\cite{ISC15,LJR16,RFD17}. 
Alternative universal gate sets, such as $\{H, CCZ\}$, can also be incorporated; 
it is known that these can be realized on hypergraph states~\cite{MM16}. 
A closely related one is the topological platform based on 
Majorana zero modes~\cite{SFN15}.
Braiding of anyons can also be realized by measurements,
but in a distinct way~\cite{BFN08}. 
We show that this can lead to topological transistors 
with enhanced fault tolerance.

This work contains the following parts.
In Sec.~\ref{sec:qtran} we explain the scheme to define 
quantum transistors and how to perform universal quantum computing with them.
In Sec.~\ref{sec:clu} we explain how to realize quantum transistors with cluster states,
and in Sec.~\ref{sec:coup} how to realize the coupling among transistors.
In Sec.~\ref{sec:phase} we extend to cluster phases and show that the transistors 
are protected by symmetry. 
In Sec.~\ref{sec:ext} we discuss how to realize transistors with 
other SPT phases, how to realize quantum diodes, and how to realize transistors in real physical systems.
We study two types of coding schemes in Sec.~\ref{sec:code} 
in order to enhance fault-tolerance. 
We conclude in Sec.~\ref{sec:conc} for further development.

\section{Quantum transistors and computation}
\label{sec:qtran}

\begin{figure}[t!]
    \centering
    \includegraphics[width=0.35\textwidth]{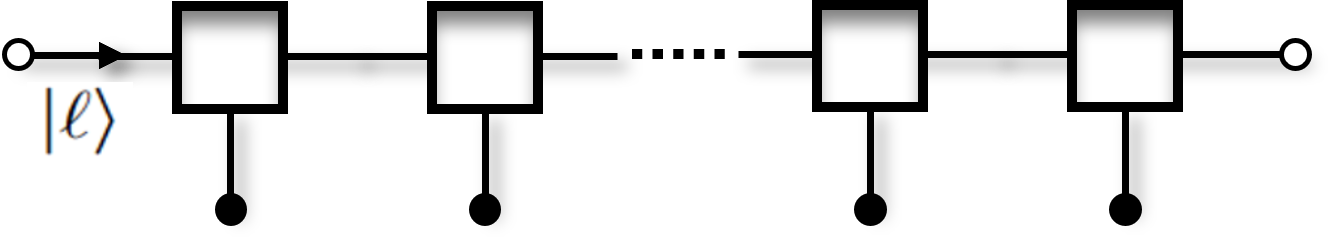}
    \caption{A quantum transistor in the form of matrix-product state with external edge modes.
    The input is on the left edge mode,
    and will be driven by measurements in the bulk and output on the right edge mode.
    The black dots are bulk sites and white dots are the two edge sites. 
    }
    \label{fig:trans}
\end{figure}

Here we introduce quantum transistors and how to perform universal quantum computing with them. 
We describe the physical systems that support quantum transistors,
and study the required control and measurement schemes on them. 
The main system we consider is the class of cluster states,
while other systems with relevant similar features are also viable. 
These systems support quantum teleportation that enables quantum gates,
and such systems are closely related to SPT order.

A quantum transistor contains three ports, 
similar with its classical counterpart:
here they are a bulk port and two edge ports.
The edge ports are for input and output,
and the bulk port is for the execution of a gate. 
The bulk system can be of various spatial dimensions, 
while we assume the edge ports are of zero dimension.
In other words, the edge ports can be viewed as quantum dots (QDs).
A proper mathematical framework to describe such systems is MPS.
In order to see this, we employ the MPS form of state
\be |\psi\ket =\sum_{i_1\cdots i_n} \Tr(E A^{i_n}\cdots A^{i_1}) |i_1 \cdots i_n\ket, \ee 
for an edge operator $E$ usually as 
$E= |\ell \ket \bra r|$,
which can be viewed as a left mode $|\ell \ket$ and a right mode $|r\ket$.

The actual (``logical'') information is carried by the edge mode,
and we need to ``fix'' the boundary condition $|r\ket$ by coupling the last site 
to an additional qudit forming a maximally entangled state (MES),
$|\omega\ket \propto \sum_{j} |jj\ket$,
also known as a qudit version of the ebit.
The first site is treated as the input mode, and its information is driven by MBQC to the last site. 
This can be viewed as an encoding with state
\be |\psi(\ell)\ket =\sum_{i_1\cdots i_n}  A^{i_n}\cdots A^{i_1} |\ell \ket |i_1 \cdots i_n\ket. 
\label{eq:qd}\ee 
With the input port, the left edge state $ |\ell \ket$ can also be fixed by forming MES,
and we use the state 
\be |\Psi\ket =\sum_{j,i_1\cdots i_n} |j \ket_L A^{i_n}\cdots A^{i_1} |j \ket_R |i_1 \cdots i_n\ket_B. 
\label{eq:choi}\ee 
The input arrives at the left port L, either by a measurement or transferred from previous steps,
and the transistor is enacted by measurement on the bulk B, 
and the output will arrive at the right port R, see Fig.~\ref{fig:trans}.
This state $|\Psi\ket $ is actually the Choi state $(V\otimes \I)|\omega\ket$
of the isometry 
\be V:=\sum_{i_1\cdots i_n} A^{i_n}\cdots A^{i_1} |i_1 \cdots i_n\ket_B, \ee 
and indeed, the transistors constructed here provide a hardware realization of the memory unit
required by the quantum von Neumann architecture~\cite{W20_choi,W22_qvn,LWLW23,W24_qvn}. 
The stored gate, which shall be unitary,
is enabled by measurement on the isometry $V$. 
This puts requirements on the local tensors $A$
that can be satisfied by some SPT ordered states.

The transistors are coupled together to form various circuits.
Types of control and measurement are needed. 
The control includes:
\begin{itemize}
    \item The execution (on/off) of a gate by a signal for bulk measurement;
    \item The possible control such as Floquet engineering to prepare a SPT system;
    \item The coupling between the QD and edge mode, and also among QDs.
\end{itemize}
The measurement includes:
\begin{itemize}
    \item The bulk measurement;
    \item The edge measurement on QD to prepare state or readout;
    \item The measurement of the bulk order parameter.
\end{itemize}

When multiple gate blocks $U_i$ are given as their Choi states $|U_i\ket$,
a compiling task is to connect them in series.
This is nontrivial if they are unknown or not Clifford gates.
Previously, a scheme based on indirect Bell measurement is developed to 
achieve this~\cite{XLS+25}, which can only guarantee correct observable on the output state,
instead of the state itself. 
For instance, for two general gates $U_1$ and $U_2$ and on input state $|\psi\ket$
the scheme can generate the correct state $|\psi_f\ket=U_2U_1|\psi\ket$,
or the shifted state $p \pi - (1-p) \psi_f$, for $\pi$ as the completely mixed state,
and $p$ as a probability parameter. 

For the transistor-based platform, 
a Choi state $|U\ket$ is a network of transistors. 
To connect two such networks, it merely requires 
the knowledge of the transistors at the interface of the two networks,
and then applies the coupling interactions to establish ebits. 
In other words, only partial knowledge of each $|U\ket$ is sufficient 
to achieve gate compiling or program composition.
This is a great advantage over qubit-based platform where gates are performed temporally, 
since the byproduct from Bell measurements cannot be corrected
for general gates.
Given a circuit to be implemented, 
the transistor-based architecture could be much faster by 
first chopping the circuit into a few blocks $U_n\cdots U_2 U_1$,
and constructing each block \emph{in parallel},
and then applying the connections among them and finally readout.
The saving in time is due to a higher cost of space.
On the contrary, 
the qubit-based architecture has to implement the gates \emph{in series}. 

We consider the requirement of fault-tolerance on the gates being considered.
We use systems with SPT order to protect the quantum transistors.
A novel feature of SPT order is the protected edge modes,
which are used to encode logical qubits at the edge ports. 
A system needs to be large enough to ensure the decoupling between the edge ports, e.g.,
larger than twice of the correlation length.
We show that symmetry protects a logical subspace within the whole edge space
and a discrete set of gates on it, which yields Clifford circuits.
Universality is achieved by extending the gate set  
with the $T$ gate by the magic-state injection scheme. 

We find that, despite the SPT order, 
the transistors are not fully robust against noises. 
The fault-tolerance can be enhanced by one additional level of software-type encoding,
or by hardware-type encoding.
Using anyons to carry qubits are a novel example of the latter.
In the following, 
we will explain the details of each piece of our construction. 

\section{Cluster states and gates}
\label{sec:clu}

In this section, we explain how to construct quantum transistors from cluster states.
We analyze the qubit case which can be easily adapted to the qudit cases.
A $n$-qubit 1D cluster state is defined as 
\be |\Phi\ket = \prod_i CZ_{i} |+\ket^n \label{eq:cluster}
\ee 
for $|+\ket^n:= |+\ket^{\otimes n}$, 
$|+\ket=\frac{1}{\sqrt{2}}(|0\ket+|1\ket)$,
and $CZ_i$ as the $CZ$ gate applied on qubit $i$ and $i+1$. 
There is an issue of ``boundary condition'': 
it is periodic (PBC) if there is one $CZ$ 
gate between the 1st and last qubit, and open (OBC) if not.
Such a cluster state is a stabilizer state with stabilizers of the form 
$Z_{i-1}X_i Z_{i+1}$ in the ``bulk'', and for OBC,
the ``edge'' stabilizers are $X_1 Z_{2}$ and $Z_{n-1} X_n$.
Denote stabilizers as $S_i$ and a Hamiltonian 
\be H=-\sum_i S_i \label{eq:h}\ee
can be defined so that its ground state is a cluster state. 

A notable feature is that there is a global $Z_2 \times Z_2$ symmetry of the model,
while its representation depends on the system.
The two symmetry operators are:
\begin{itemize}
    \item 
$\otimes_{i\in even} X_i$ and $\otimes_{i\in odd} X_i$ for PBC and even $n$;
\item $Z_1\otimes_{i\in even} X_i Z_n$ and $Z_1\otimes_{i\in odd} X_iZ_n$ for PBC and odd $n$;
\item $Z_1 \otimes_{i\in even} X_i$ and $\otimes_{i\in odd} X_i Z_n$ for OBC and even $n$;
\item $Z_1 \otimes_{i\in even} X_i Z_n$ and $\otimes_{i\in odd} X_i$ for OBC and odd $n$.
\end{itemize}

No matter which form,
the symmetry of the Hamiltonian is also preserved by the state,
so it has symmetry-protected topological (SPT) order, 
and it is represented projectively on the effective edge mode. 
In order to see its symmetry in the MPS formalism,
we first note that the cluster state~(\ref{eq:cluster}) is a Choi state in the form~(\ref{eq:choi})
and the first (last) site can be identified as the input (output) site.
The state is 
translation-invariant, 
so it is only specified by two 
operators for each site, which are $A_0=HP_0 /\sqrt{2}$, $A_1=HP_1 /\sqrt{2}$
for $H$ as the Hadamard gate, $P_0=|0\ket \bra 0|$ and $P_1=|1\ket \bra 1|$
are projectors.
In the Pauli X basis, the tensor becomes $B_i=HZ^i$ ($i=0,1$).
In order to see the action of the symmetry, we have to group two sites
(see Fig.~\ref{fig:trans}) and 
\be B_{00}=\I/2,  B_{01}=X/2, B_{10}=Z/2, B_{11}=XZ/2. \ee 
Using the general symmetry condition 
\be \sum_j u_{ij} M_j= V^\dagger M_i V \ee
for unitary operators $U=(u_{ij})$ and $V$,
or equivalently, $(U\otimes \I)\hat{M}= V^\dagger \hat{M} V$
for $\hat{M}:=\sum_i |i\ket \otimes M_i$,
the symmetry condition for the cluster tensor, denoted as $\hat{B}$, are
\be (Z\otimes \I)\hat{B} = X \hat{B}X,\; 
(\I \otimes Z)\hat{B} = Z \hat{B}Z. \ee 
The projective representation, as the defining feature of SPT order, 
is seen from the Pauli relation $XZ=-ZX$.

Next we describe how to realize $H$ and $S$ gates with 1D cluster state,
and then explain how to realize $CZ$ gate with two-leg ladder cluster states and extensions.
In MBQC, one applies on-site projective measurement and record the outcome.
For the 1D case, we need to use OBC and choose one end (usually the left) as the input, 
and the other end as the output.
The first site is assumed to carry a more general initial state $|\psi\ket$,
and the edge term $X_1Z_2$ is no longer a stabilizer. 
The underlying mechanism is quantum teleportation,
and the usual teleportation is understood as 2-bit teleportation
based on 1-bit teleportation~\cite{ZLC00}.
Given a state $|\psi \ket$ and an ancilla $|+\ket$, 
the 1-bit teleportation is 
\be \bra s|(H\otimes \I) CZ |\psi\ket |+\ket=HZ^s|\psi\ket \ee
for $s=0,1$ as the projective measurement outcome.
It is then clear to see two-step of such processes leads to 
$X^tZ^s |\psi\ket $.

A sequence of $n$ measurements in X basis leads to the gate sequence 
\be HZ^{i_{n-1}} \cdots HZ^{i_2} HZ^{i_1},  \ee 
and it can be written as 
 \be Z^{\sum_{even} i_k} X^{\sum_{odd} i_k} H \ee
up to a nonphysical $\pm 1$ phase factor if $n$ is even.
This means the total parity on the even (odd) sites contributes a $Z$ ($X$) byproduct,
so it is convenient to treat the 1D system as a bipartite lattice.
Different from the usual MBQC which requires on-site measurement,
here we only need the global measurement for the two sublattices.
This could be a substantial improvement for realization requirement.
It is essential that the total number of measured sites is odd.
If it is even, then the final gate is an identity gate
as Pauli gates are usually treated as byproduct operators
or assumed to be easy to realize. 
This difference can be used to calibrate a transistor on the first hand before being used.

Usually, $S$ gate and even any Z-basis rotation can be realized by one on-site measurement 
due to the 1-bit teleportation.
Here instead we use a global scheme. 
Applying a transversal $S^{\otimes n}$ 
on the cluster state before the X-basis measurement,
it is easy to find the outcome gate sequence is
\be  QQZ^{i_{n-1}} \cdots QZ^{i_2} Q Z^{i_1} \ee
for $Q:=HS$. 
It is interesting to note $Q^3=\I$ up to a phase, 
and the byproduct in the final outcome gate sequence 
collects into three parts $X^{\sum_{k=1}^{\lfloor \frac{n-1}{3} \rfloor} i_{3k-2}}$,
$Z^{\sum_{k=1}^{\lfloor \frac{n-2}{3} \rfloor} i_{3k-1}}$, 
$Y^{\sum_{k=1}^{\lfloor \frac{n}{3} \rfloor} i_{3k}}$, 
in which $\lfloor \dots \rfloor$ is the floor function.
So for a measurement length multiple of three, we get the $Q$ gate. 
The phase gate $S$ switches $X$ and $Y$, 
so the state before the measurement is still a stabilizer state.
We can also observe a phenomena of period on gates, and this
is a character of quantum cellular automaton (QCA)~\cite{Arr19}.
It has been explored for cluster states but for different tasks~\cite{Rau05a,Rau05b}. 

The gate set $\{H,Q\}$ is universal for qubit-Clifford gate compiling.
As we consider fault-tolerant scheme in this work,
we restrict the gates being discrete and elementary.
The initialization of the edge mode is also a nontrivial fact,
so we only assume the initial state being $|0\ket$ or $|+\ket$ as in a normal cluster state.

\begin{figure}[t!]
    \centering
    \includegraphics[width=0.3\textwidth]{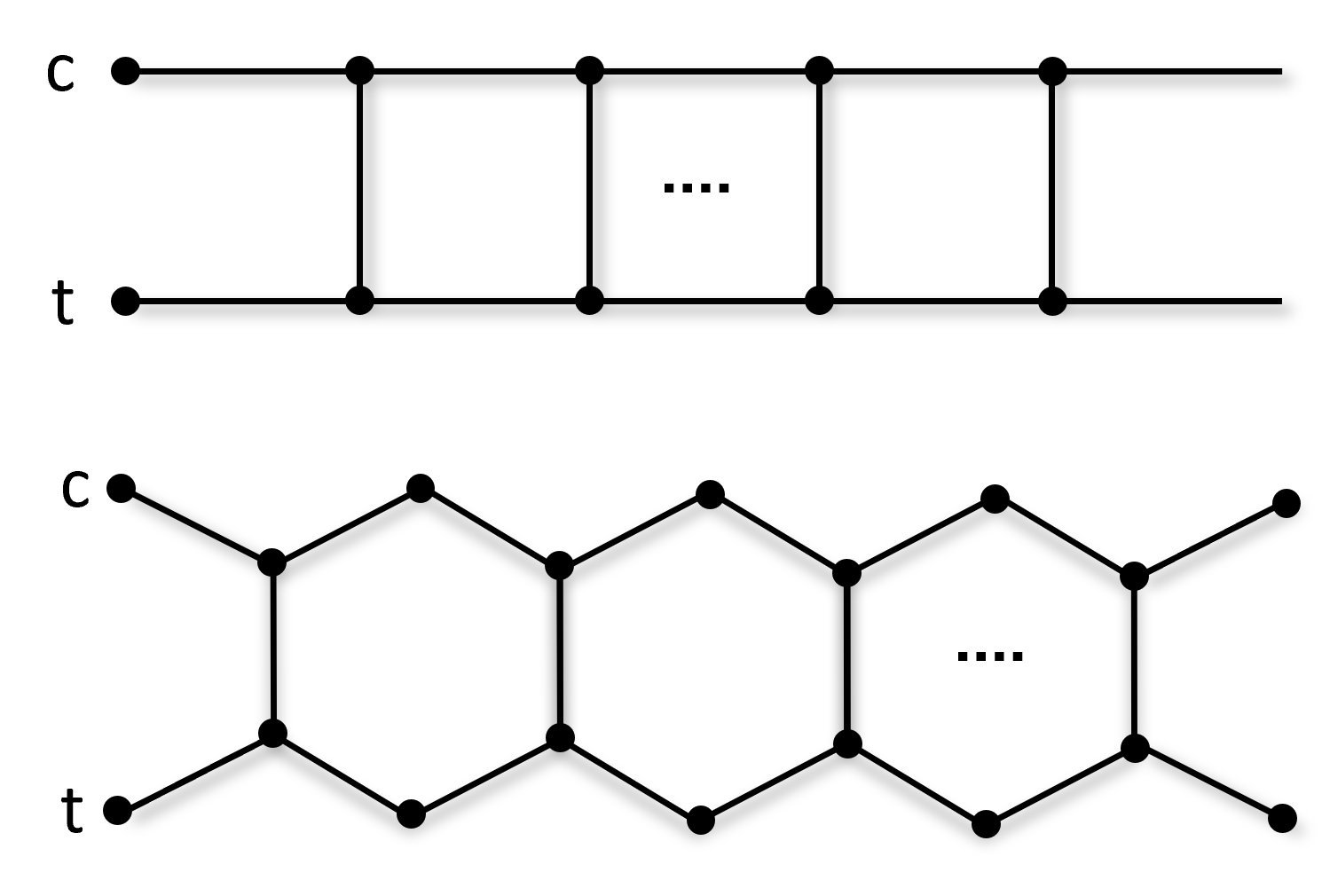}
    \caption{Two-leg ladders to realize $CZ$ gate. 
    Top: square lattice; Bottom: honeycomb lattice.
    The control (c) and target (t) qubits are on the left of the cluster.
    Time evolves towards the right. 
    }
    \label{fig:cluster2}
\end{figure}

Now we consider the entangling gate $CZ$. 
We find two schemes that use two types of cluster states.
They are still based on the teleportation schemes above. 
The 1-bit scheme realizes $CZ(H\otimes H)$ and leads to 
a cluster state on a two-leg ladder of square lattice.
The 2-bit scheme realizes $CZ$ and leads to 
a cluster state on a two-leg ladder of honeycomb lattice,
see Fig.~\ref{fig:cluster2}.
Although the 2-bit scheme has a higher cost, 
it realizes $CZ$ directly. 
The gate $CZ (H \otimes H)$ is of order 6, 
and its cube is a swap gate. 

The byproduct operator can also be easily obtained. 
For instance, for the honeycomb lattice
we measure in the X basis, and 
the total byproduct operator is of the form 
\be Z^{\sum_{odd} a_o +\sum_{even} b_e} X^{\sum_{even} a_e} \otimes 
Z^{\sum_{odd} b_o +\sum_{even} a_e} X^{\sum_{even} b_e}
\ee 
for $a$, $b$ as outcome of each leg.

Furthermore, we find 
the width of the ladder can be extended. 
In order to see this, 
we present a symmetry-based method.
First, we can use symmetry to explain the scheme above to realize 
$H$ and $S$ gates.
For the $H$ gate, the symmetry is seen from the list below Eq.~(\ref{eq:h}).
By measuring $X$ on all sites except the 1st and last sites,
the symmetry operator 
$Z_1 \otimes_{i\in even} X_i$
reduces to $Z_1 X_n$. 
This is the stabilizer for the entangled state of the 1st and last sites,
and a measurement on the 1st site will teleport the information
to the last site. 
This logic applies to the $S$ gate and also the $CZ$ gate. 

Actually, cluster states and their Hamiltonian on 2D lattices are known to show 
peculiar symmetries. 
On the square lattice, a cluster state has line-like and cone-like 
subsystem symmetry~\cite{MM16,ROW+19,SNB+19,DAM20},
derived from products of stabilizers.
On the honeycomb lattice, a cluster state has fractal subsystem symmetry~\cite{DAM20}.
This can be used to encode many qubits by treating the open modes on the left
as input, and on the right as output.
The two-leg ladders are special cases:
each of the two edges (top and bottom) still have $Z_2\times Z_2$ symmetry,
but with different representations from the 1D cluster chain.
In this work, for each block of 2D cluster state,
we only consider one control mode and one target mode. 
We will name the two cluster states as square cluster 
and honeycomb cluster, for simplicity.

For the square cluster, a symmetry operator would
take the form $E_l X_b E_r$,
for $E_{l,r}$ denoting operators acting on the edges,
and $X_b$ denoting strings of Pauli $X$s acting in the bulk.
Measuring those $X$s would yield $E_l E_r$ as a stabilizer.
Similar with the 1D case, there is a size effect:
the width $w$ and length $\ell$ of a cluster are related.
Define \be T_w=(\otimes_j CZ_j)(\otimes_j H_j) \label{eq:tw} \ee 
for $j\in [1,w]$ running through
a column and $CZ_j$ acting on a nearest pair of sites $(j,j+1)$ in it.
A global $X$ measurement in the bulk will lead to $T_w^\ell$, 
and a seminal result is that $T_w^{w+1}$ is a reflection operator~\cite{Rau05b},
known as Raussendorf's reflection (RR). 
That is, it swaps the 1st row and the last row, and so on.

Now we explain how to realize a logical $CZ$ gate. 
Similar with the 1D case, we use the bulk to propagate information
while the edge to execute the $CZ$ gate from $T_w$.
For a width $w$, the length needs to be $\ell=2(w+1)m+1$ 
since $T_w^{2(w+1)}=\I$, for $m\in \mathbb{Z}$.
Let the control and target be on the 1st and last row, 
and assume the initial state $|\alpha\ket |0\cdots 0\ket |\beta\ket$,
one application of $T_w$ actually leads to a cluster state 
with edges $|\alpha\ket$ and $|\beta\ket$ along a column.
Now it becomes easy to see measuring the sites in the middle 
can lead to a $CZ$ gate on $|\alpha\ket |\beta\ket$.
This is also known as the remote $CZ$ scheme~\cite{ZLC00}. 
This has a requirement on the width to be $w=2n$
since pairs of ebits need to be formed in between the 1st and last row. 

For the honeycomb cluster, it works similarly except that now 
a symmetry operator is of fractal form, known as Sierpinski triangle~\cite{DAM20,SNB+19}.
Its transfer operator $T_w$ is a two-step QCA 
\be T_w=(\otimes_{r\in even} U_{r,r+1})(\otimes_{r\in odd} U_{r,r+1}) \ee
for $U_{r,r+1}:=CZ(H\otimes H)$, 
and it also generates a cluster chain which finally yields the desired $CZ$ gate on the edges.
However, in this case 
it is not easy to find the period of a fractal symmetry operator,
and sometimes it may grow fast.
It is known that~\cite{SNB+19} for width $w=2^k$, the period is also $2^k$,
for $k\in \mathbb{Z}$, so
the length needs to be $\ell=mw+1$, for $m\in \mathbb{Z}$.
Note here the length counts the steps of the $T_w$ gate. 
Our construction of logical $CZ$ gate can also be extended to other lattices, 
while the square lattice is the simplest case. 

\section{Coupling and gate compiling}
\label{sec:coup}

\begin{figure}[t!]
    \centering
    \includegraphics[width=0.45\textwidth]{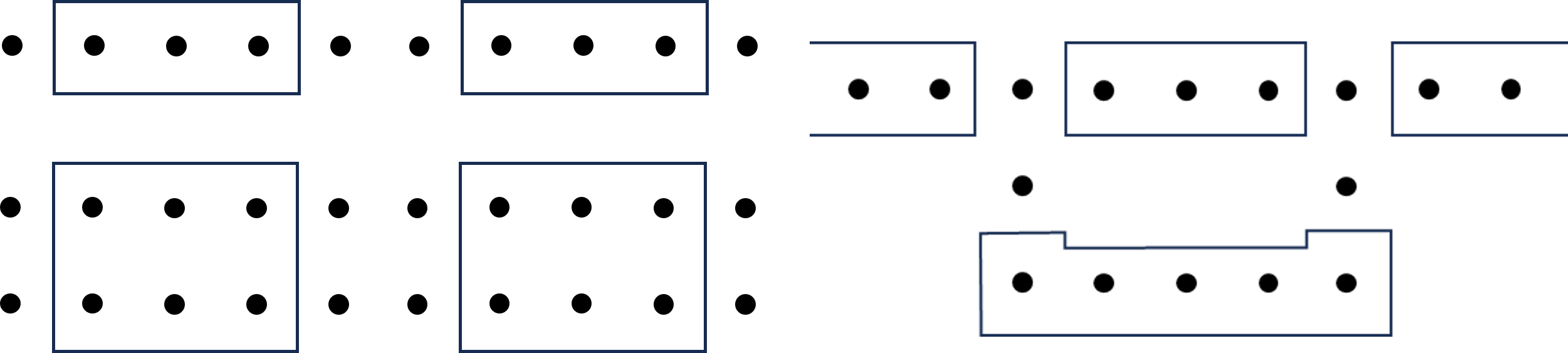}
    \caption{Coupling among gates. 
    Top left: coupling between qubit gates; 
    Bottom left: coupling between $CZ$ gates;
    Right: conditional gate. 
    A box represents a bulk system. 
    The chosen number of dots are only for illustration.
    }
    \label{fig:cluster3}
\end{figure}

The $T$ gate is needed to complete a universal gate set. 
It is the square root of the phase gate $S$.  
It is also realized via teleportation, 
which is usually known as magic-state injection. 
It is realized as 
\be T|\psi\ket= (P_i\otimes S^i) CX |t\ket |\psi\ket \ee 
for the magic state $|t\ket=T|+\ket$, 
and controlled-not gate $CX$ with $|t\ket$ as the target,
and the byproduct correction is a phase gate $S$, for $i=0,1$.
Note the conjugate of $T$ can also be realized by a slight modification of it.
The gate $CX$ is equivalent to $CZ$ from the relation $HZH=X$.

We see that in order to realize $T$, 
a sequence of $H$, $CZ$, and conditional $S$ gates are needed. 
This requires the coupling among the gates. 
The primary scheme is to use a qubit to couple to an edge mode
forming an ebit, so that a state can be teleported from the qubit to the edge mode,
or vice versa. 
The external qubit serves as an effective edge mode 
and can be carried by a quantum dot or other systems, 
or even extended to a bulk system, e.g., a chain of cluster state. 
It plays the role of ``wire'' for the free evolution ($\I$ gate) of qubits.

The basic structures for the couplings are shown in Fig.~\ref{fig:cluster3}.
The mechanism is to change the edge terms of the cluster stabilizers 
into bulk terms.
It is easy to find these terms.
For instance, 
for the 1D H-H coupling, 
it modifies the original edge term $ZX$ to $ZXZ$, $XZ$ to $ZXZ$,
and for H-S coupling, 
it modifies the original edge term $ZX$ to $ZXZ$, $YZ$ to $ZYZ$.
For the junction between two $CZ$ gates, 
four qubits are sufficient.
The control and target ports are separately controlled.

To realize the conditional $S$ gate required by the $T$ gate,
multiple-qubit couplings may be needed.
One such configuration is shown in the right panel of Fig.~\ref{fig:cluster3},
and weight-4 stabilizer is needed at the junction. 
When one gate is selected, 
the other gate can be decoupled via Z-basis measurement.

In all, we see that a network of gates forms a giant cluster, 
with magic states on the edges. 
The Pauli byproduct from the measurements for each transistor is corrected `on-site'
without the need to propagate it. 
In addition, one can also take a global viewpoint of the scheme above, 
by preparing a cluster on the first hand, 
and then identify the parts for each gate and junction,
and also the magic states. 
However, when we extend the scheme to cluster phases, 
it is preferable to take the local modular viewpoint 
as each gate is locally protected by a SPT phase. 

\section{Cluster phases and gates}
\label{sec:phase}

Cluster states are known as fixed points of SPT phases.
It has been expected that there are universal features of a SPT phase 
that can be used for quantum computing~\cite{ESB+12},
and it has been explored under the paradigm 
of universal phases of quantum matter~\cite{Wei18}.
Here we shift the focus from universality to fault-tolerance,
and we take a modular point of view and 
study the fault-tolerance of each gate. 
Compared to previous schemes, 
our scheme of realizing gates is more succinct.

To highlight the difference, 
we briefly recall the previous scheme~\cite{RWP17,SWP+17,ROW+19}.
For a properly defined cluster phase containing the cluster state, 
a ground state in the phase is no longer a stabilizer state.
In the MPS form, the bond dimension gets larger. 
Indeed, 
it has been shown that any translation-invariant 
state in a 1D bosonic $Z_2\times Z_2$ SPT phase 
is specified by tensor of the form 
\be B_i = \sigma_i \otimes J_i \label{eq:tensorb}
\ee 
for $\sigma_i$ as Pauli operators, 
and the edge space is in a maximally non-commutative (MNC)
cohomology class~\cite{ESB+12}.
In general, there are direct-sum blocks in the tensors
if the MNC condition is dropped, but 
each block is of the same bipartite structure~\cite{Ste17}.
The part $\sigma_i$ is protected by the symmetry, 
and the part $J_i$, with indefinite size, 
carries the details for each state 
which can be viewed as the disturbance from the cluster point. 
The corresponding spaces are denoted
as the logical space $\C H_L$ and junk space $\C H_J$, respectively. 
Here the index $i$ defines a physical basis that will be referred to 
as the ``wire'' basis. 
There is also a choice of basis for the edge space, 
and this can be fixed by initialization of input state. 

The scheme developed earlier~\cite{RWP17,SWP+17,ROW+19,SNB+19,DAM20,Wei18,RYA23} 
used measurements away from the wire basis. 
This would mix the logical and junk operators,
leading to non-unitary operation on the logical system.
It hence restricted to tiny angles to approximate unitary gates,
and used a so-called ``oblivious wire'' in between any two gates 
to avoid the accumulation of disturbance.
The oblivious wire used the wire basis measurement on a segment,
but it would correct the logical Pauli byproduct while ignoring 
the outcomes for the junk part.

\begin{figure}[t!]
    \centering
    \includegraphics[width=0.45\textwidth]{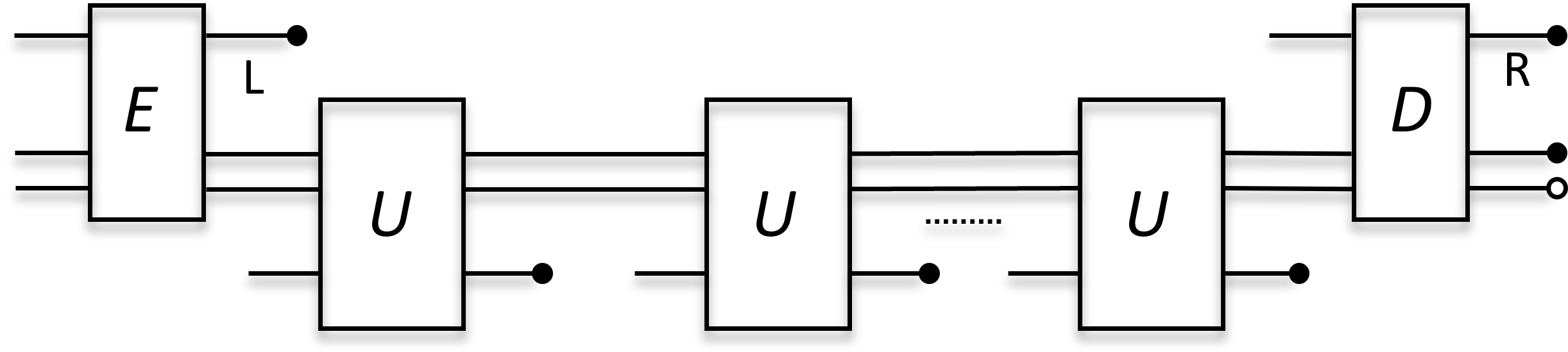}
    \caption{The encoding structure for a ground state in a cluster phase.
    The double-line is the bond space,
    the black dots are qubits, and the small circle is the ``junk space.''
    }
    \label{fig:cluster4}
\end{figure}

Our scheme develop in this work utilizes a central feature 
that is omitted previously. 
Namely, the wire basis measurement of odd length on 1D cluster state 
can realize the Hadamard gate $H$,
compared with the identity gate $\I$ for even length.
We have shown above the $S$ gate is realized similarly,
hence we can avoid the oblivious-wire approach. 
We will argue below our $H$ and $S$ gates, also $CZ$,
are of SPT nature.

\subsection{Encoding structure}

First, we analyze the encoding structure. 
For the special point of cluster state,
it can be viewed as an encoding $|\ell\ket \mapsto |\psi(\ell)\ket$. 
It is a SPT code and has a constant code distance~\cite{WZO+20}.
Its logical operators are near the edge:
the logical $X_L$ is $XZ$, and logical $Z_L$ is just $Z$.
It is a stabilizer code and the MBQC on the chain can be viewed as a
one-way code switching, 
which gradually deletes the stabilizers from the left until the right end.
When extended into a phase,
the code distance gets larger but still remains constant.
In other words, an edge mode now is a finite-width wavepacket 
but its correlation with the bulk decays exponentially.

Regarding fault-tolerance,
the nontrivial issue here is that the edge state $|\ell\ket$
can be an entangled state on $\C H_L \otimes \C H_J$. 
In order to understand this, we show in Fig.~\ref{fig:cluster4}
the encoding structure, for the left QD qubit labeled L,
and right one by R.
They are each coupled to the edge mode by a unitary operator $E$ and $D$,
respectively.
The unitary operator $U$ in the bulk realizes the tensor $B_i$ for each site.
The whole state takes the form 
\be |\Phi\ket = \sum_{\vec{i}} D B_{\vec{i}} E |\vec{i}\ket |+\ket_L |+\ket_R,  \ee 
for both L and R initialized at $|+\ket$,
$B_{\vec{i}}$ as the product of bulk tensors, and 
$\vec{i}$ as bulk sites. 
The operator $E$ ($D$) encodes (decodes) information in the bond space.
The state above shall be a Choi state in the form~(\ref{eq:choi})
for perfect gate execution.
The cluster state is a special case:
there is no junk space, 
and the logical space is played by the 2nd qubit near the left edge,
and $E$ and $D$ both is $CZ$ gate.

\subsection{Single-site tensor}

We see from~(\ref{eq:tensorb}) there are only Pauli operators on the logical space, 
hence the measurement in the wire basis can only realize free evolution, 
modular Pauli gates. 
Note that the wire basis is the X basis referring to the cluster chain.
This is distinct from our scheme above which uses $H$ or $S$ in the tensor.
In order to see how the gate $H$ or $S$ is protected by the SPT order, 
we need to understand better the form~(\ref{eq:tensorb}).
It is derived in Ref.~\cite{ESB+12} via a group-theoretic approach.
It implicitly assumed a translation-invariant representation $u(g)$ 
for the group element $g\in G$,
and for the 1D cluster state, this requires an even length.
It needs to group two sites together to make the representation of $Z_2\times Z_2$
on-site and translation-invariant: as $X\I$ and $\I X$.

However, we find that by allowing edge anomaly and non-translation-invariant representation
\be U(g)=\otimes_r u_r(g), \ee 
we can still apply the same method~\cite{ESB+12}.
In this case, $u_r(g)$ and $u_r(h)$ may not commute for site $r$ near the edge
for $g,h\in G$ and $[g,h]=0$,
but $[U(g),U(h)]=0$, and $[u_r(g),u_r(h)]=0$ for site $r$ away from the edge.
This is used to define the wire basis $\{|s\ket\}$
so that any $u_r(g)$ is diagonal.
The site-dependent action on the tensor is 
\be \sum_{s'} u_r(g)_{ss'}  A^{[r]}_{s'} = e^{i\theta_r(g)} 
W_r(g)^\dagger A^{[r]}_s W_{r-1}(g). \ee
The projective representation $W_r(g)$ at different site corresponds 
to the same cocycle of $G$.
For the MNC setting, $W_r(g) \cong V(g) \otimes \I_{junk}$.
There must exist basis transformation operator $G_r$ with 
\be G_r^{\dagger} W_r(g) G_r = V(g) \otimes \I_{junk},\ee
and similarly for $W_{r-1}(g)$.
For $G=Z_2\times Z_2$, 
$V(g)$ are Pauli operators acting on the logical space and will be denoted as $\sigma^{[r]}_s$.
We find the single-site tensor as 
\be A^{[r]}_s = G_r (\sigma^{[r]}_s \otimes J^{[r]}_s) G_{r-1}^{\dagger}.\ee
Note here the local physical dimension for $s$ is two,
instead of four.
The operator $G_n$ on the last site $n$ may not be the inverse of $G_1$ on the first site.
If we block two sites of $A^{[r]}_s$,
it is still a tensor-product form, which takes 
the form~(\ref{eq:tensorb}) as the translation-invariant special case. 

To reduce to the cluster-state point,
we can use a bipartite lattice: a tensor $\{\I,X\}$ and $\{\I,Z\}$
for two sites. As $X=HZH$, one $H$ gate can be bundled to a tensor. 
The $H$ or $S$ gate can also be extracted to act on the bond, 
treated as a local gauge operator.
This explains the realization of $H$ and $S$ gates
since qubit Clifford group is generated by them.

\subsection{Symmetry protection}

We now analyze the symmetry protection for $H$ and $S$ gates.
Actually, this is quite obvious from our symmetry-based method:
as long as the $Z_2\times Z_2$ symmetry is preserved,
measuring strings of $X$s in the bulk will lead to an ebit between the edges.
For the $H$ gate, the two symmetry operators are of the form
$X\I X\I\cdots X$, $ZX\I X\I X\cdots \I X Z$.
They do not commute locally on the two edges.
Measuring bulk $X$s leads to $XX$ and $ZZ$ defining an ebit.
That is to say, the $H$ as well as $S$ gates are symmetry-protected.
The reason is the same for the identity gate.

Different from the identity gate, however,
the requirement of symmetry protection here is more restrictive.
Due to the edge anomaly, 
no additional terms beyond the stabilizers are allowed.
This is to `pin' the edge modes.
On the contrary, there can be additional bulk terms such as on-site $X$.
Using the encoding picture in Fig.~\ref{fig:cluster4},
modular the isometry $V$ in the middle, 
the encoding operator $E$ will generate an entangled state 
in the form $\sum_\ell |\ell\ket_L |\phi_\ell\ket$,
for $|\phi_\ell\ket\in \C H_L\otimes \C H_J$,
while the decoding operator $D$ will map $|\phi_\ell\ket$
onto $|\ell\ket_R$, 
yielding the ebit state $\sum_\ell  |\ell\ket_L|\ell\ket_R$.

A formalism based on string order instead of MPS was developed recently~\cite{RYA23}.
A string order starting on site $r$ is the expectation value of
\be O_r(g)=v_r(g) \otimes \left[\otimes_{j=r+1}^{n-1} u_j(g) \right]\otimes v_{n}(g) \ee 
for $v_r(g)$, $v_{n}(g)$ as projective representation,
$u_j(g)$ as linear representation, $g\in G$.
This includes symmetry operators as the special case $U(g)=O_1(g)$.
The oblivious-wire approach~\cite{RWP17,SWP+17,ROW+19} 
extracts continuous rotation gates from the string operators.
However, our approach extracts $H$ gate from $U(g)$,
and similarly $S$ gate from a corresponding symmetry representation.

The requirement for pinning the edges might be challenging in practice.
If the symmetry is violated on the edges,
the $H$ and $S$ gates are not perfect anymore,
and the `remote' ebit between the two edges will become noisy.
In order to analyze the effect of noise,
we assume we start from a 1D wire of even length for the identity gate,
yet we only performed an odd number of $X$ measurements.
This will lead to an operator $G$ as a perturbation of $H\otimes \I_{junk}$ 
acting on the bond space.
The operator $G$ does not factorize into two parts. 
This means that any nontrivial logical gate cannot be exactly realized 
by a simple wire-basis measurement. 

Suppose the perturbation parameter is $\lambda \ll 1$ and module $H$ gate,
the unitary operator $G$ can be expanded to the linear approximation 
\be G(\lambda) \approx (\I + \lambda g), \ee
in which $g$ is an anti-Hermitian operator,
and the on-site tensor can be approximated as 
\be A_s \approx (\I + \lambda g) 
\left[ \sigma_s \otimes (\I + \lambda \tilde{A}_s) \right]. \ee
Ignoring the term $O(\lambda^2)$, we find 
\be A_s-  (\sigma_s \otimes \I) \approx 
\lambda \left[(\sigma_s \otimes \tilde{A}_s)+ g (\sigma_s \otimes \I) \right]\in O(\lambda).
\ee 
We see that the infidelity and the symmetry-violating 
perturbation parameter $\lambda$ are linearly related in the first order.
At this point, one may wonder that if one more site is measured,
a perfect identity gate can be realized.
This is because there will be one additional gauge operator $G^\dagger$ on the bond space
that cancels the effect of $G$.

Overall, the goal can be understood as 
to extract an ebit between L and R by the measurement in the bulk.
The bulk, L, and R will be in an entangled state, 
and a further measurement on the bulk will generate an entangled state $|\varphi\ket_{LR}$.
The fidelity between $|\varphi\ket_{LR}$ and the ebit $|\omega\ket$ is upper bounded by 1,
depending on the junk space, or equivalently, the string order.
The pinning of edge condition enforces the symmetry, leading to the ebit.
Given ebits, remote $CZ$ or $CX$ gate can be realized by the gate teleportation scheme.

The relation between gate fidelity and string order was recognized before~\cite{DB09}.
The recent work~\cite{RYA23} showed that string order and logical operator values
can be obtained by local measurements. 
This leads to a scheme to prompt the fidelity.
Namely, one can first make a 
measurement of its logical values of $Z_L$ or $X_L$, 
and project onto an eigenstate of them before the coupling to the external edge QD systems.
This constitutes a scheme for initialization. 
After it, the effective edge mode is a qubit.
It is then similar to use an edge Hamiltonian to couple the external QD to the edge mode,
forming a bond between them.

\subsection{CZ gate}

We now analyze the symmetry-protection of logical $CZ$ gates.
The oblivious-wire approach has been applied to the case of 2D square cluster,
by treating it effectively as the 1D case~\cite{ROW+19}.
Namely, by grouping a $w\times w$ square block of sites 
(compared with two sites for 1D), it leads to a SPT order for the group 
$Z_2^w\otimes Z_2^w$.
The method of Ref.~\cite{ESB+12} then applies leading to a similar form of its local tensors
as~(\ref{eq:tensorb}).
On the contrary, in our formalism there is no need to block sites along the 
``time direction,'' so only $w$ qubits are blocked along the space direction,
and this can encodes $w$ logical qubits.
However, we would only use the top and bottom ones for the control and target qubits.

In order to see the SPT nature, 
we shall analyze the symmetry $Z_2^w\otimes Z_2^w$.
A symmetry operator is deduced from a product of stabilizers.
It turns out a $Z_2$ symmetry operator can be of cone shape or 
line-like~\cite{ROW+19,SNB+19,DAM20}, both being a rigid subsystem symmetry.
By treating 2D square cluster as coupled 1D cluster wires,
we can also choose it to be the 1D symmetry along the horizontal direction, 
with only a different representation.
For instance, a symmetry operator $X\I X\I \cdots X\I$ now becomes 
\bea \nonumber
&Z\;\I \;Z\;\I \cdots Z\;\I \\
&X\;\I \;X\;\I \cdots X \; \I \\ \nonumber
&Z\;\I \;Z\;\I \cdots Z\;\I 
\eea 
that acts on three neighboring wires in the bulk.
This is due to the action of arrays of $CZ$ gates coupling the wires.

Instead of a Hadamard gate $H$, now the element in a QCA is the $T_w$ gate~(\ref{eq:tw}).
As long as there is no symmetry-breaking near the edges,
this gate is SPT.
Our argument also applies to the honeycomb cluster.
Therefore, the logical $CZ$ gate is protected by symmetry. 

We arrived at the symmetry-protection without invoking the tensor-network framework. 
This is possible due to the symmetry constraint.
The MPS and tensor network are a `picture' that can show the encoding structure.
Tensor-network or PEPS forms of cluster states have been studied~\cite{ROW+19},
which, however, do not help directly for our study 
since in our scheme we have to pick one space direction as the simulated time direction.
Using the coupled-wire picture, 
we can use the 1D tensor and the easily verified fact that 
$CZ$ acting on the two physical qubits can be converted into a $CZ$ acting the two bond qubits,
so that the tensor for two vertical pair of sites is 
\be A_{ij}= CZ(H\otimes H)(Z^i\otimes Z^j). \ee 
This extends to the $T_w$ gate~(\ref{eq:tw}) for a larger width.

The product of the subsystem symmetry operators also lead to a 
global $Z_2$ symmetry, and will have a $Z_2$ SPT order. 
However, it turns out its SPT order from the 3rd cohomology $\C H^3(Z_2,\mathbb{C})$ is trivial.
The nontrivial one corresponds to a hypergraph state
that uses the $CCZ$ gate instead of $CZ$ for preparation.
The $CCZ$ contributes to a nontrivial 3-cocycle.
This can also be understood from the gauging mechanism~\cite{LG12},
with the cluster phase mapped to toric code, 
and hypergraph phase mapped to double-semion model.

\section{Extensions}\label{sec:ext}

In this section, we discuss extensions of the cluster-phase based transistor scheme above.
It can be extended to other SPT phases, and even to topological (TOP) phases.
As the classical case, 
there also exists the two-port quantum diodes as reduction of transistors. 

Recall that our scheme differs from traditional schemes in MBQC.
We do not assume a whole substrate and local adaptive measurements on it;
instead, we use modular pieces of substrate, i.e., our transistors,
and global measurements on each of them,
plus controllable couplings among them.
Our scheme highlights the role of symmetry-protection for a transistor,
with the bulk mainly responsible for transfer/teleportation,
and the edge mainly responsible for the gate.
This picture also carries over to other SPT phases.

\subsection{SPT phases}

We mainly consider two classes of other SPT phases.
The 2D cluster phase is an example of subsystem SPT order  
with nontrivial 
2-cocycle but trivial 3-cocycle based on cohomology.
It is a so-called ``weak'' SPT order,
and another seminal example is the valence-bond solids (VBS)~\cite{AKLT87},
which have global continuous symmetry.
An example of ``strong'' SPT order is the class of 2D
hypergraph states (HGS) and their phases. 
Both VBS and HGS have been shown as resources for 
MBQC~\cite{GE07,BM08,WAR11,Miy11,W19_rev,MM16,GGM19,HSF24,Tak24}.

The construction of VBS applies to a broad range of spin systems
with Lie symmetry, such as the orthogonal, unitary, or symplectic group 
being useful for MBQC~\cite{WSR17}.
Here we consider the usual $SO(3)$ case. 
A seminal system is 
the 1D VBS model of spin-1 chain defined by a frustration-free
two-body Hamiltonian $H=\sum_n H_n$, for $H_n$
as $P_2$, the projector onto spin-2 sector.
It is a special point for the following model
\be H= \sum_n \vec{S}_n \cdot\vec{S}_{n+1} - \beta (\vec{S}_n \cdot\vec{S}_{n+1})^2 \ee 
at $\beta=-1/3$, belong to the so-called Haldane phase~\cite{AL86} for $|\beta|<1$.
Actually, a Haldane phase only requires the $Z_2\times Z_2$ subgroup of the full $SO(3)$ symmetry,
or the anti-unitary time-reversal symmetry~\cite{PBT+12}.
Here we require the full SO(3) symmetry,
and as such, we take the basis of spin-z operator as the wire basis.

To construct a transistor, we attach two spin-$1/2$, i.e., qubit,
at its two edges. 
Edge interactions between a spin-$1/2$ and a spin-$1$ 
have been studied before~\cite{Miy10,BBM10} respecting the SO(3) symmetry,
and a symmetry representation takes the form
\be U(g)=v^*(g)\otimes\left[\otimes_r u(g) \right]\otimes v(g) \ee 
for any $g\in SO(3)$, and $u(g)$ as linear representation, 
and $v(g)$ as projective representation.
It is clear from the symmetry, the edge part $v^*(g)\otimes v(g)$
defines the ebit as its unique representation.
Therefore, once a state $|\psi\ket$ arrives at the left edge,
a logical gate $v(g)$ can be implemented by measuring the bulk and the right edge 
in the rotated basis defined by $U(g)$,
with Pauli byproduct corrected according to the measurement outcomes.
Given the coupling between two edges, constructed via the VBS formalism,
the state $v(g)|\psi\ket$ is hence teleported to the next transistor. 
Also note a gate $v(g)$ is realized deterministically since 
it originates from the symmetry operation, 
instead of linear combination of the Pauli byproducts as previously employed.

For entangling gates, we shall use 2D VBS that have been well studied for MBQC.
We refer to the review~\cite{Wei18} for the details. 
The spin-$3/2$ VBS on the honeycomb lattice is a seminal example,
and it has been shown the $CX$ gate can be realized directly from 
two neighboring sites~\cite{Miy11}.
Then from similar scheme with the 2D honeycomb cluster states, 
we can realize the $CZ$ as well as $CX$ transistors. 
This also has a requirement on the system size,
and note it may not directly extend to VBS on other lattices.
Coupling between the edges of 1D VBS and 2D VBS can be easily constructed.

We shall comment on the differences from previous schemes for MBQC with VBS, 
e.g., in Refs.~\cite{Miy10,BBM10,WSR17}.
(i) We do not require a single phase that needs to be universal.
Although a 1D VBS wire can be `cut out' from a 2D VBS, 
that is unnecessary.
(ii) Different from 1D cluster phase, here one can obtain any qubit gate in SU(2) 
directly, without using the discrete set of $H$, $S$, and $T$.
This leads to a saving of space but more restrictive requirement of symmetry protection.
Even if the full symmetry is realized, 
one can still only apply the discrete gates $H$ and $T$ 
to enhance fault-tolerance since realizing an arbitrary rotation angle directly 
may not be robust.
(iii) The logical gates are realized on ``edges'' that are protected by the 
SPT order of the whole system, hence there is no need to 
go `deep' in the bulk to perform gates.

Finally, we briefly study the class of HGS. 
Instead of the $CZ$ gate, it uses $CCZ$ or even higher-order control
versions to define a HGS.
It is known that GHS can realize teleportation,
hence can be universal for MBQC~\cite{W19_rev,MM16,GGM19,HSF24,Tak24}.
The HGS defined on 2D lattices can show 2D strong SPT order of the global symmetry $Z_2$,
and can realize different universal gate set, 
such as the set of $H$ and $CCZ$.
However, the byproduct operator can be the $CZ$ gate, 
which is a nontrivial gate to implement on its own.
The $CZ$ gate can be realized by a cluster transistor,
or simulated by $CCZ$ gate with a control qubit set as value 1.
A 1D chain of hypergraph state can be viewed as an encoded version of cluster state, 
and posses $Z_2 \times Z_2$ SPT order~\cite{HSF24},
hence can be used to realize $H$ gate.
There is no need to realize the $T$ gate anymore.
However, the 2D lattices to realize the logical $CCZ$ gate are more involved.
Also, if treated as ground states,
the parent Hamiltonian will be more complicated.

\subsection{Quantum diodes}

Besides transistors, one may wonder how to construct quantum diodes. 
It turns out this is straightforward,
and this also implies that our construction of quantum transistors is a natural framework. 

A quantum diode is obtained by deleting one edge port of a quantum transistor. 
Actually, the state of a quantum diode can be written as Eq.~(\ref{eq:qd}),
and can be viewed 
as a state of a covariant code~\cite{WZO+20,WWC+22,ZLJ20,YMR+22}.
It is an encoding of $|\ell\ket$ as $|\psi(\ell)\ket$ which has a symmetry 
that can directly convert the gate on the former to a transversal gate on the latter.
One can easily verify that it has the `unidirectional' feature:
the measurement on the bulk can prepare a logical state on the edge, 
but not the other way. 
Furthermore, a quantum diode can be connected to a quantum transistor 
by connecting their edges, 
but the `null-edge' port of a diode cannot be connected to a transistor. 

It is interesting to compare our scheme of quantum diodes 
to previous schemes. 
We use measurement on the bulk to induce logical gates,
which is also a decoding scheme.
In the standard covariant code scheme logical gates would be 
realized by transversal unitary operations based on symmetry.
Most of the known MBQC schemes also only specify the input edge,
without an explicit construction of the output edge 
or `pushing it' far away as it is irrelevant for the universality analysis.

\subsection{Physical realization}

In this section, we analyze the physical realization of 
quantum transistors in actual systems. 
We focus on the cluster case, 
and other SPT order can be realized in a similar way.
Depending on the current technology, 
there are mainly the following ways:
\begin{enumerate}
    \item Realized in actual quantum materials;
    \item Simulated in controllable quantum computing platforms
    or prepared by quantum circuits;
    \item Generated via Floquet engineering on 
    actual quantum materials or simulated quantum systems.
\end{enumerate}

The first type mostly requires the realization of the parent
Hamiltonian.  
Due to the many-body nature of cluster stabilizers,
it is well known that 
it is not easy to realize a parent cluster Hamiltonian 
with two-body spin interactions~\cite{GB08}.
This is different from VBS,
which can be realized in spin-1 chain or spin-$1/2$ ladders in molecular chains 
or materials~\cite{HKA+90}.
It is not clear, though, if measurement on the bulk can be easily done 
to distill ebit pair between two edge modes.

It is straightforward to prepare a cluster state by quantum circuit~\cite{CWC2023,OKA2025}. 
This can be realized by current quantum simulators.
In this setting, the states are not ground states,
hence there is no passive protection by a Hamiltonian, hence by symmetry.
A noisy cluster state may not be contained by a well-defined cluster phase.
Despite this, one still can use coding technique to 
ensure fault tolerance.

A distinct approach is the Floquet engineering~\cite{BCO+09},
which uses high-frequency driving to induce effective interaction  
based on the Magnus expansion.
This has been studied in the setting of quantum magnetism~\cite{ISC15,LJR16,RFD17}.
A Floquet 1D cluster phase can be 
generated by the following model 
\be H(t)=h\sum_i X_i + \Theta(t) f(t)\sum_i Z_i Z_{i+1} \ee 
for the driving profile $f(t)=\lambda \omega \cos(\omega t+\varphi)$ $(\lambda >0)$
and Heaviside $\Theta(t)$ step function~\cite{ISC15}.
It also has $Z_2\times Z_2$ symmetry, 
yet one of the symmetry sector is an emergent one in the time direction,
and it is only approximate w.r.t. the driving frequency $\omega$.
This phase is separated from other phases 
such as the paramagnetic (PM) or ferromagnetic (FM) ones.
It is also possible to generate 2D cluster phase by Floquet Ising interactions.

The Floquet scheme is appealing to realize our quantum transistors. 
Before the computation, a transistor could be in a usual classical phase,
PM or FM.
Then the Floquet driving is turned on to generate the stabilizer terms and
the edge modes. 
Measurements will execute both state transfer and gates, 
but collapse the devices. 
The collapsed state is a product state belonging to a PM or FM phase,
so it would be straightforward to drive it back to the cluster phase by 
turning on the Floquet driving again. 

Furthermore, it has been found that adding disorder can induce many-body localized 
phases preserving the symmetry not only at the ground states~\cite{KLM+16,FWV+22}.
This can suppress thermal noises hence increase the coherence time. 
These techniques are promising yet these proposals 
are so far only realized by quantum simulators~\cite{PPS+17,KDP18,MCD+21}.
We expect that these proposals can also be realized in bulk quantum materials~\cite{OK19}.

\section{Coding}\label{sec:code}

In order to enhance fault-tolerance, 
we can employ coding scheme based on transistor network.
Here we discuss two such strategies: 
one is to employ a higher-level coding by treating the transistors as building blocks,
and the other is to employ a lower-level coding by constructing more robust encoded transistors.
Apparently, these two strategies can also be combined together.

\subsection{Stabilizer codes}

Stabilizer codes allow coding with good parameters~\cite{Got98}.
A quantum code is usually denoted as $[[n,k,d]]$ for encoding $k$ logical qubits
into $n$ physical qubits with code distance $d$.
Both $k$ and $d$ can scale with $n$ for good codes such as 
low-density parity check (LDPC) codes~\cite{BE21}. 
A code can be prepared by implementing its isometric encoding operator $V$,
or by measuring its stabilizers. 
Another method is that any stabilizer state in a code 
can be expressed as a graph state~\cite{CM18},
hence one can prepare its simple logical state, e.g., $|0\ket_L$,
as a graph state by $H$ and $CZ$ gates.
A decoding can be realized by $V^\dagger$ followed by computational basis measurement
on ancilla, or by measuring its stabilizers.
On current qubit-based architecture,
a common approach to realize coding is to measure its stabilizers
for both the encoding and decoding operations.

To realize coding for our transistor-based architecture,
we can use either the measurement-based or isometry $V$ for either the encoding or decoding.
It would depend on the coding structure to choose one scheme in practice. 
Here we discuss two issues that are different from the qubit-based architecture. 

There can be long-range stabilizers or gates that need to permute qubits around.
This can lead to crossing of propagation `wires' to shuffle qubits,
which is also a problem that is central to electrical circuit~\cite{HH13}. 
To avoid this, we can use swap and reflection module based on the gate~(\ref{eq:tw}).
This is different from current platforms, e.g., 
for atomic systems an atom can be moved by lasers or external fields~\cite{BEG+24,ABB+24},
and for solid-state systems such as superconducting circuits a qubit is moved by 
implementing teleportation or swap gates,
while for photonic systems photons can move quite freely.

Based on the modular feature, we can realize the encoding $E$, logical gates $U_i$,
and decoding $D$ as separate blocks.
We know there are Clifford and non-Clifford type gate blocks,
and the encoding/decoding can be of measurement or unitary type.
If the refreshing cycle is used,      
a fault-tolerant gate sequence $DU_n \cdots U_2DU_1E$
can be realized as a quantum sequential circuit~\cite{W26},
with each piece of block can be reused.
After the quantum decoding operation,
a classical decoder algorithm is also needed to determine the optimal error recovery
by Pauli operators.
The fusion between gate blocks needs to be performed after the classical decoder.

In addition, there are also other types of coding,
such as entanglement-assisted codes~\cite{LB13} that use free entanglement
between the encoder and decoder, 
convolutional codes~\cite{FP14} that perform encoding at various time stage,
and program codes~\cite{LW25} that use Choi-preserving superchannels as encoding. 
All these codes are consistent with the transistor architecture,
and we leave them for further study.

\begin{figure}
    \centering
    \includegraphics[width=0.45\textwidth]{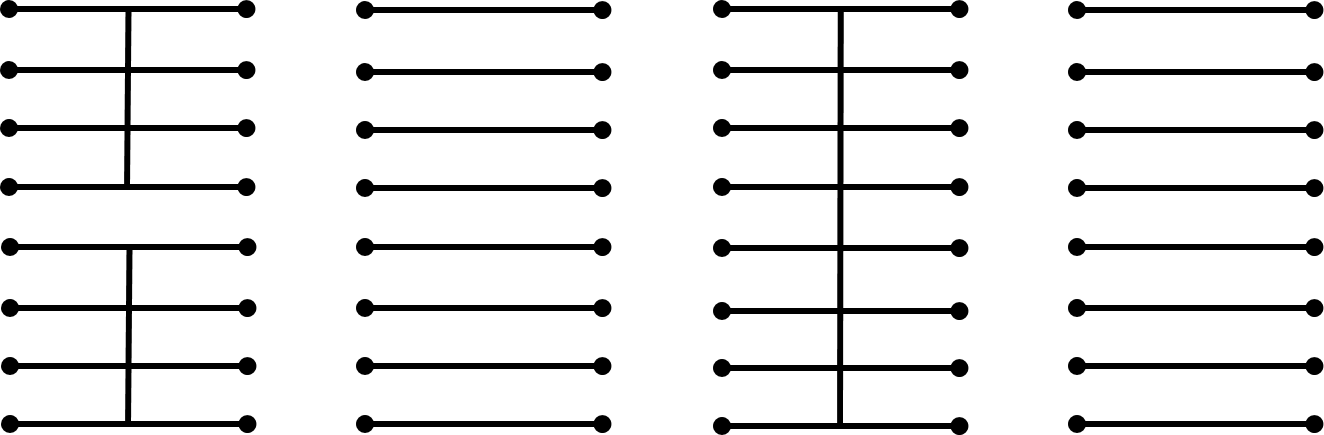}
    \caption{An illustration of the quantum computer architecture with 
    topological transistors. The horizontal direction is evolution time.
    Each wire with two end points is an anyonic ebit, 
    and a group of four of them with a connecting vertical wire 
    is a topological transistor, 
    with each end point as an anyon
    (here referring to Majorana fermion).
    The width-eight array supports the entangling $CX$ or $CZ$ gate.
    Other ebits are used to realize anyonic teleportation via parity measurement.
    Qubit Clifford gates can be realized by braiding the four anyons in one column of a 
    transistor.
    }
    \label{fig:toptr}
\end{figure}

\subsection{Topological substrates}

Our construction of quantum transistors relies on SPT order. 
One may wonder if the usual topological (TOP) order~\cite{ZCZ+15} 
can be employed to define transistors.
A first glance reveals some similarity between 1D SPT wires 
and topological wires, such as the Majorana wire~\cite{Kit01,SFN15,ZLW19,Mic25}. 
The idea of MBQC has also been used to enable anyon braiding~\cite{BFN08,ZDJ16}
for topological quantum computing (TQC)~\cite{NSS+08}.
It turns out the similarity is not valid, 
yet still we can modify the usual setup to construct topological transistors.
From coding perspective,
this constitutes a concatenation by using a lower-level TOP codes.

Different from SPT order, 
the logical space from TOP order is the degenerate fusion space of anyons.
It takes a direct-sum form $\C H=\oplus_a \C H_a$ for $a$ labeling anyon types,
and it corresponds to the direct-sum block form of local tensors
in the PEPS form of a TOP state~\cite{SCP10}.
Anyon braidings are unitary operators acting on the whole space $\C H$,
while the logical actions do not depend on the local details of each sector $\C H_a$.

It is easy to see a TOP state $|\Psi(a_1,a_2,\dots)\ket$ 
with a set of anyons is of the form of a quantum diode,
with the logical state of the anyons encoded in the whole system.
In order to extend it to quantum transistors, 
we need to construct ebits and Choi states of anyons.
Anyonic ebits and teleportation have been defined~\cite{SFN15},
with a pair of anyons $(\gamma_1,\gamma_2)$ with a fixed parity $\gamma_1\gamma_2$
as anyonic ebit, 
and then a sequence of parity measurements can teleport/transfer 
an anyon $\gamma_0$ to another anyon $\gamma_3$.

We now describe a scheme for Majorana fermions which are non-Abelian anyons.
A qubit can be encoded in the fixed parity subspace 
$\gamma_1\gamma_2\gamma_3\gamma_4=1$ of four anyons.
Then with eight anyons, shown in Fig.~\ref{fig:toptr}, 
we can define a logical anyonic ebit,
with each horizontal pair in a fixed parity sector,
and each vertical quartet also in a fixed parity sector.
Next, we need to see how to \emph{store} the braiding operations 
in anyonic ebits,
and it turns out there is no need to design particular structure,
as braiding is realized by a sequence of parity measurements~\cite{Mic25}.
The $CZ$ or $CX$ gate requires an array of anyons of width eight,
with each four for one qubit.
As braiding of Majorana fermions realizes Clifford gates, 
similar with cluster phase, 
one needs other gates to achieve universality. 
A common method is to realize the $T$ gate by magic-state injection,
which is not fully topological. 

The TOP transistors would be more robust against local noises than the SPT transistors.
On the other hand, the parity measurements on anyons,
e.g., by dispersive coupling to external quantum dots~\cite{Mic25},
could be more challenging than the on-site measurements on SPT edge modes.
Our scheme also differs from the current TQC.
We employs the space-time map to `store' the braiding gates 
at various time by more anyons. 
As such, our scheme has a higher space cost, but as the trade-off,
each anyon or anyonic qubit only needs to survive for a shorter time period
since its state would be teleported to the next station.
Also we shall recall the difference between 
the roles of measurements for TQC and MBQC.
For the former, the parity measurements, which appear to be local
but actually are nonlocal, 
are used to induce anyonic teleportation, 
which is used to simulate braiding operations that form logical gates.
For the latter, global measurements are used 
to induce gate teleportation which realizes logical gates directly.

\section{Conclusion}\label{sec:conc}

\begin{table}[]
    \centering\footnotesize
    \renewcommand{\cellalign}{cc}
    \begin{tabular}{|c|c|c|c|c|}
        \hline
        & \bf{\makecell{Transistor-\\ based}} & \bf{\makecell{Qubit-\\ based}} & \bf{\makecell{Fusion-\\ based}} & \bf{MPQW} \\
        \hline
        Elements & transistors & qubits & ebits & \makecell{local\\ excitations} \\
        \hline
        Coupling & \makecell{$P_\omega$ via\\ local H terms} & non & \makecell{$P_\omega$ via \\ 2-photon \\ fusion } & \makecell{local H terms} \\
        \hline
        Gates & NEM & \makecell{1\&2-qubit \\ gates in SU(2)} & NEM & \makecell{local evolution\\ $e^{itH}$} \\
        \hline
        C/M, I/O & NEM & NEM & NEM & NEM \\
        \hline
    \end{tabular}
    \caption{A brief comparison of hardware details for a few universal quantum computing architecture,
    including our transistor-based, the usual qubit-based, photonic fusion-based~\cite{BBB21}, 
    and multiparticle quantum walk (MPQW)~\cite{CGW13}. 
    C/M: control and measurements; I/O: input and output;
    NEM: non-entangling measurements; 
     $P_\omega$: projection on ebits.}
    \label{tab:comp2}
\end{table}

In this work, we studied universal quantum computing architecture 
based on our construction of quantum transistors. 
A quantum transistor stores a quantum gate that can be executed via 
measurement-induced gate teleportation. 
We can name a teleportation-based quantum transistor in particular as ``telesistor'',
since, in principle, there could also be other types of mechanism to 
construct quantum transistors. 

To further understand the features of our architecture, 
we can compare it with a few others shown in Table~\ref{tab:comp2}.
We see that it uses local interaction terms to connect transistors, 
effectively realizing projections onto Bell states, or known as ebits. 
This is similar with the fusion gate in photonic fusion-based model~\cite{BBB21},
which is used to construct graph states~\cite{RB01}, 
but differs from the local coupling in the Hamiltonian-based model such as 
multiparticle quantum walk (MPQW)~\cite{CGW13}.
For the qubit-based model, although storage of states by an array of qubits is simple, 
realizing gates are more complicated requiring coupling to external lasers or fields.

In abstract terms, the qubit-based circuit model is universal,
hence can be used to simulate any other universal architecture. 
A transistor in the form of Choi state can be simulated by qubits.
That said, the von Neumann architecture can be realized solely via qubits. 
What makes the transistors unique shall be due to hardware features:
bulk materials that enable transistors can have symmetry protection,
the bulk part of transistors can also be used to carry information and acted upon 
by further quantum operations~\cite{W26}. 
Besides, it also allows integration of both qubits and gates 
to construct large-scale integrated quantum chips.

What remains to be seen is what materials can be used to construct the transistors. 
Although our gates are protected by symmetry, 
the symmetry itself could be broken in real system. 
There are attempts to restore the symmetry-protection to thermal or noisy regimes~\cite{MW23,DZLY25}, 
whether this could work, especially under Floquet control, is an open question.
From the development of digital computers based on magnetic cores, vacuum tubes, or transistors,
advantages must be found for quantum transistors 
in order to make them indispensable.

\section{Acknowledgement}

This work has been funded by
the National Natural Science Foundation of China under Grants
12447101 and 12105343 (Y.-D.L., X.X., D.-S.W.), and 12274250 (Q.-R.W.).

\end{spacing}

\bibliography{ext}{}

\begin{thebibliography}{10}
\expandafter\ifx\csname url\endcsname\relax
  \def\url#1{\texttt{#1}}\fi
\expandafter\ifx\csname urlprefix\endcsname\relax\def\urlprefix{URL }\fi
\expandafter\ifx\csname href\endcsname\relax
  \def\href#1#2{#2} \def\path#1{#1}\fi

\bibitem{NC00}
M.~A. Nielsen, I.~L. Chuang, Quantum Computation and Quantum Information,
  Cambridge University Press, Cambridge U.K., 2000.

\bibitem{LJL+10}
T.~D. Ladd, F.~Jelezko, R.~Laflamme, Y.~Nakamura, C.~Monroe, J.~L. O’Brien,
  Quantum computers, Nature 464~(7285) (2010) 45--53.

\bibitem{RB01}
R.~Raussendorf, H.~J. Briegel, A one-way quantum computer, Phys. Rev. Lett. 86
  (2001) 5188--5191.

\bibitem{BK05}
S.~Bravyi, A.~Kitaev, Universal quantum computation with ideal clifford gates
  and noisy ancillas, Phys. Rev. A 71 (2005) 022316.

\bibitem{CGW13}
A.~M. Childs, D.~Gosset, Z.~Webb, Universal computation by multiparticle
  quantum walk, Science 339 (2013) 791.

\bibitem{CMP18}
T.~S. Cubitt, A.~Montanaro, S.~Piddock, Universal quantum hamiltonians,
  Proceedings of the National Academy of Sciences 115~(38) (2018) 9497--9502.

\bibitem{AL18}
T.~Albash, D.~A. Lidar, Adiabatic quantum computation, Rev. Mod. Phys. 90
  (2018) 015002.

\bibitem{Arr19}
P.~Arrighi, An overview of quantum cellular automata, Natural Computing 18
  (2019) 885--899.

\bibitem{YRC20}
Y.~Yang, R.~Renner, G.~Chiribella, Optimal universal programming of unitary
  gates, Phys. Rev. Lett. 125 (2020) 210501.

\bibitem{W24rev}
D.-S. Wang, Universal quantum computing models: a perspective of resource
  theory, Acta Phys. Sin. 73 (2024) 220302.

\bibitem{HH13}
D.~M. Harris, S.~L. Harris, Digital design and computer architecture, Elsevier,
  2013.

\bibitem{W22_qvn}
D.-S. Wang, {A prototype of quantum von Neumann architecture}, Commun. Theor.
  Phys. 74 (2022) 095103.

\bibitem{Cho75}
M.-D. Choi, Completely positive linear maps on complex matrices, Linear Algebra
  Appl. 10 (1975) 285--290.

\bibitem{W26}
D.-S. Wang, Quantum sequential circuits, arXiv preprint arXiv:2602.05166
  (2026).

\bibitem{BBC+93}
C.~H. Bennett, G.~Brassard, C.~Cr\'epeau, R.~Jozsa, A.~Peres, W.~K. Wootters,
  Teleporting an unknown quantum state via dual classical and
  einstein-podolsky-rosen channels, Phys. Rev. Lett. 70 (1993) 1895--1899.

\bibitem{GC99}
D.~Gottesman, I.~L. Chuang, Demonstrating the viability of universal quantum
  computation using teleportation and single-qubit operations, Nature
  402~(6760) (1999) 390--393.

\bibitem{ZLC00}
X.~Zhou, D.~W. Leung, I.~L. Chuang, Methodology for quantum logic gate
  construction, Phys. Rev. A 62 (2000) 052316.

\bibitem{note_bit}
We note that the classical case is more nuanced; for instance, bits can also be
  stored in hardware such as hard drives.

\bibitem{KLM01}
E.~Knill, R.~Laflamme, G.~Milburn, A scheme for efficient quantum computation
  with linear optics, Nature (London) 409 (2001) 46.

\bibitem{BFC13}
D.~Bacon, S.~T. Flammia, G.~M. Crosswhite, Adiabatic quantum transistors, Phys.
  Rev. X 3 (2013) 021015.

\bibitem{WB15}
D.~J. Williamson, S.~D. Bartlett, Symmetry-protected adiabatic quantum
  transistors, New J. Phys. 17 (2015) 053019.

\bibitem{AK25}
M.~Asoudeh, V.~Karimipour, A review of perfect quantum state transfer, from one
  to two and three dimensional arrays of qubits, International J. Theor. Phys.
  64 (2025) 198.

\bibitem{WSR17}
D.-S. Wang, D.~T. Stephen, R.~Raussendorf, Qudit quantum computation on matrix
  product states with global symmetry, Phys. Rev. A 95 (2017) 032312.

\bibitem{RWP17}
R.~Raussendorf, D.-S. Wang, A.~Prakash, T.-C. Wei, D.~T. Stephen,
  Symmetry-protected topological phases with uniform computational power in one
  dimension, Phys. Rev. A 96 (2017) 012302.

\bibitem{SWP+17}
D.~T. Stephen, D.-S. Wang, A.~Prakash, T.-C. Wei, R.~Raussendorf, Computational
  power of symmetry-protected topological phases, Phys. Rev. Lett. 119 (2017)
  010504.

\bibitem{ROW+19}
R.~Raussendorf, C.~Okay, D.-S. Wang, D.~T. Stephen, H.~P. Nautrup,
  Computationally universal phase of quantum matter, Phys. Rev. Lett. 122
  (2019) 090501.

\bibitem{SNB+19}
D.~T. Stephen, H.~P. Nautrup, J.~Bermejo-Vega, J.~Eisert, R.~Raussendorf,
  Subsystem symmetries, quantum cellular automata, and computational phases of
  quantum matter, Quantum 3 (2019) 142.

\bibitem{DAM20}
A.~K. Daniel, R.~N. Alexander, A.~Miyake, Computational universality of
  symmetry-protected topologically ordered cluster phases on 2{D} {A}rchimedean
  lattices, {Quantum} 4 (2020) 228.

\bibitem{Wei18}
T.-C. Wei, Symmetry-protected phases for measurement-based quantum computation,
  Advances in Physics: X 3 (2018) 1461026.

\bibitem{RYA23}
R.~Raussendorf, W.~Yang, A.~Adhikary, Measurement-based quantum computation in
  finite one-dimensional systems: string order implies computational power,
  {Quantum} 7 (2023) 1215.

\bibitem{GW09}
Z.-C. Gu, X.-G. Wen, Tensor-entanglement-filtering renormalization approach and
  symmetry-protected topological order, Phys. Rev. B 80 (2009) 155131.

\bibitem{BM08}
G.~K. Brennen, A.~Miyake, Measurement-based quantum computer in the gapped
  ground state of a two-body hamiltonian, Phys. Rev. Lett. 101 (2008) 010502.

\bibitem{DB09}
A.~C. Doherty, S.~D. Bartlett, Identifying phases of quantum many-body systems
  that are universal for quantum computation, Phys. Rev. Lett. 103 (2009)
  020506.

\bibitem{BBM10}
S.~D. Bartlett, G.~K. Brennen, A.~Miyake, J.~M. Renes, Quantum computational
  renormalization in the haldane phase, Phys. Rev. Lett. 105 (2010) 110502.

\bibitem{Miy10}
A.~Miyake, Quantum computation on the edge of a symmetry-protected topological
  order, Phys. Rev. Lett. 105 (2010) 040501.

\bibitem{RMB13}
J.~M. Renes, A.~Miyake, G.~K. Brennen, S.~D. Bartlett, Holonomic quantum
  computing in symmetry-protected ground states of spin chains, New J. Phys. 15
  (2013) 025020.

\bibitem{HSF24}
Y.~Hong, D.~T. Stephen, A.~J. Friedman, Quantum teleportation implies
  symmetry-protected topological order, Quantum 8 (2024) 1499.

\bibitem{WZO+20}
D.-S. Wang, G.~Zhu, C.~Okay, R.~Laflamme, Quasi-exact quantum computation,
  Phys. Rev. Res. 2 (2020) 033116.

\bibitem{WWC+22}
D.-S. Wang, Y.-J. Wang, N.~Cao, B.~Zeng, R.~Laflamme, Theory of quasi-exact
  fault-tolerant quantum computing and valence-bond-solid codes, New J. Phys.
  24 (2022) 023019.

\bibitem{ZLJ20}
S.~Zhou, Z.-W. Liu, L.~Jiang, New perspectives on covariant quantum error
  correction, Quantum 5 (2021) 521.

\bibitem{YMR+22}
Y.~Yang, Y.~Mo, J.~M. Renes, G.~Chiribella, M.~P. Woods, Optimal universal
  quantum error correction via bounded reference frames, Phys. Rev. Res. 4
  (2022) 023107.

\bibitem{PVW+07}
D.~Perez-Garcia, F.~Verstraete, M.~M. Wolf, J.~I. Cirac, Matrix product state
  representations, Quantum Inf. Comput. 7~(5) (2007) 401--430.

\bibitem{ESB+12}
D.~V. Else, I.~Schwarz, S.~D. Bartlett, A.~C. Doherty, Symmetry-protected
  phases for measurement-based quantum computation, Phys. Rev. Lett. 108 (2012)
  240505.

\bibitem{AKLT87}
I.~Affleck, T.~Kennedy, E.~H. Lieb, H.~Tasaki, Rigorous results on valence-bond
  ground states in antiferromagnets, Phys. Rev. Lett. 59 (1987) 799--802.

\bibitem{GPB+22}
S.~Galeski, K.~Y. Povarov, D.~Blosser, S.~Gvasaliya, R.~Wawrzynczak,
  J.~Ollivier, J.~Gooth, A.~Zheludev, {$LT$ Scaling in Depleted Quantum Spin
  Ladders}, Phys. Rev. Lett. 128 (2022) 237201.

\bibitem{PEC+24}
J.~Philippe, F.~Elson, N.~P.~M. Casati, S.~Sanz, M.~Metzelaars, O.~Shliakhtun,
  O.~K. Forslund, J.~Lass, T.~Shiroka, A.~Linden, D.~G. Mazzone, J.~Ollivier,
  S.~Shin, M.~Medarde, B.~Lake, M.~M\aa{}nsson, M.~Bartkowiak, B.~Normand,
  P.~K\"ogerler, Y.~Sassa, M.~Janoschek, G.~Simutis,
  ${({\mathrm{C}}_{5}{\mathrm{H}}_{9}{\mathrm{NH}}_{3})}_{2}{\mathrm{cubr}}_{4}$:
  A metal-organic two-ladder quantum magnet, Phys. Rev. B 110 (2024) 094101.

\bibitem{JCA+25}
I.~Jakovac, T.~Cvitani\ifmmode~\acute{c}\else \'{c}\fi{},
  D.~Ar\ifmmode~\check{c}\else \v{c}\fi{}on, M.~Herak, D.~Cin\ifmmode
  \check{c}\else \v{c}\fi{}i\ifmmode~\acute{c}\else \'{c}\fi{}, N.~B.
  Topi\ifmmode~\acute{c}\else \'{c}\fi{}, Y.~Hosokoshi, T.~Ono, K.~Iwashita,
  N.~Hayashi, N.~Amaya, A.~Matsuo, K.~Kindo, I.~Lon\ifmmode \check{c}\else
  \v{c}\fi{}ari\ifmmode~\acute{c}\else \'{c}\fi{},
  M.~Horvati\ifmmode~\acute{c}\else \'{c}\fi{}, M.~Takigawa, M.~S.
  Grbi\ifmmode~\acute{c}\else \'{c}\fi{}, Properties of an organic model $s=1$
  haldane chain system, Phys. Rev. B 111 (2025) 064407.

\bibitem{ERA25}
C.~Edmunds, E.~Rico, I.~Arrazola, G.~Brennen, M.~Meth, R.~Blatt, M.~Ringbauer,
  Symmetry-protected topological haldane phase on a qudit quantum processor,
  PRX Quantum 6 (2025) 020349.

\bibitem{ISC15}
T.~Iadecola, L.~H. Santos, C.~Chamon, Stroboscopic symmetry-protected
  topological phases, Phys. Rev. B 92 (2015) 125107.

\bibitem{LJR16}
T.~E. Lee, Y.~N. Joglekar, P.~Richerme, String order via floquet interactions
  in atomic systems, Phys. Rev. A 94 (2016) 023610.

\bibitem{RFD17}
A.~Russomanno, B.-e. Friedman, E.~G. Dalla~Torre, Spin and topological order in
  a periodically driven spin chain, Phys. Rev. B 96 (2017) 045422.

\bibitem{MM16}
J.~Miller, A.~Miyake, Hierarchy of universal entanglement in 2d
  measurement-based quantum computation, npj Quantum Information 2 (2016)
  16036.

\bibitem{SFN15}
S.~D. Sarma, M.~Freedman, C.~Nayak, Majorana zero modes and topological quantum
  computation, npj Quantum Information 1 (2015) 15001.

\bibitem{BFN08}
P.~Bonderson, M.~Freedman, C.~Nayak, Measurement-only topological quantum
  computation, Phys. Rev. Lett. 101 (2008) 010501.

\bibitem{W20_choi}
D.-S. Wang, Choi states, symmetry-based quantum gate teleportation, and
  stored-program quantum computing, Phys. Rev. A 101 (2020) 052311.

\bibitem{LWLW23}
Y.-T. Liu, K.~Wang, Y.-D. Liu, D.-S. Wang, {A Survey of Universal Quantum von
  Neumann Architecture}, Entropy 25~(8) (2023) 1187.

\bibitem{W24_qvn}
D.-S. Wang, {A family of quantum von Neumann architecture}, Chin. Phys. B 33
  (2024) 080302.

\bibitem{XLS+25}
X.~Xu, Y.-D. Liu, S.~Shi, Y.-J. Wang, D.-S. Wang, Distributed quantum computing
  with black-box subroutines, Quantum Sci. Technol. 10 (2025) 045014.

\bibitem{Rau05a}
R.~Raussendorf, Quantum cellular automaton for universal quantum computation,
  Phys. Rev. A 72 (2005) 022301.

\bibitem{Rau05b}
R.~Raussendorf, Quantum computation via translation-invariant operations on a
  chain of qubits, Phys. Rev. A 72 (2005) 052301.

\bibitem{Ste17}
D.~T. Stephen, Computational power of one-dimensional symmetry-protected
  topological phases, {Master Thesis, University of British Columbia (2017)}.

\bibitem{LG12}
M.~Levin, Z.-C. Gu, Braiding statistics approach to symmetry-protected
  topological phases, Phys. Rev. B 86 (2012) 115109.

\bibitem{GE07}
D.~Gross, J.~Eisert, Novel schemes for measurement-based quantum computation,
  Phys. Rev. Lett. 98 (2007) 220503.

\bibitem{WAR11}
T.-C. Wei, I.~Affleck, R.~Raussendorf, {Affleck-Kennedy-Lieb-Tasaki State on a
  Honeycomb Lattice is a Universal Quantum Computational Resource}, Phys. Rev.
  Lett. 106 (2011) 070501.

\bibitem{Miy11}
A.~Miyake, Quantum computational capability of a 2d valence bond solid phase,
  Annals of Physics 326~(7) (2011) 1656 -- 1671, july 2011 Special Issue.

\bibitem{W19_rev}
D.-S. Wang, Quantum computation by teleportation and symmetry, Int. J. Mod.
  Phys. B 33~(15) (2019) 1930004.

\bibitem{GGM19}
M.~Gachechiladze, O.~G\"uhne, A.~Miyake, Changing the circuit-depth complexity
  of measurement-based quantum computation with hypergraph states, Phys. Rev. A
  99 (2019) 052304.

\bibitem{Tak24}
Y.~Takeuchi, Catalytic transformation from computationally universal to
  strictly universal measurement-based quantum computation, Phys. Rev. Lett.
  133 (2024) 050601.

\bibitem{AL86}
I.~Affleck, E.~H. Lieb, A proof of part of haldane's conjecture on spin chains,
  in: Condensed Matter Physics and Exactly Soluble Models, Springer, 1986, pp.
  235--247.

\bibitem{PBT+12}
F.~Pollmann, E.~Berg, A.~M. Turner, M.~Oshikawa, Symmetry protection of
  topological phases in one-dimensional quantum spin systems, Phys. Rev. B 85
  (2012) 075125.

\bibitem{GB08}
T.~Griffin, S.~D. Bartlett, Spin lattices with two-body hamiltonians for which
  the ground state encodes a cluster state, Phys. Rev. A 78 (2008) 062306.

\bibitem{HKA+90}
M.~Hagiwara, K.~Katsumata, I.~Affleck, B.~I. Halperin, J.~P. Renard,
  Observation of s=1/2 degrees of freedom in an s=1 linear-chain heisenberg
  antiferromagnet, Phys. Rev. Lett. 65 (1990) 3181--3184.

\bibitem{CWC2023}
S.~Cao, B.~Wu, F.~Chen, M.~Gong, Y.~Wu, Y.~Ye, C.~Zha, H.~Qian, C.~Ying,
  S.~Guo, et~al., Generation of genuine entanglement up to 51 superconducting
  qubits, Nature 619~(7971) (2023) 738--742.

\bibitem{OKA2025}
J.~O’Sullivan, K.~Reuer, A.~Grigorev, X.~Dai, A.~Hern{\'a}ndez-Ant{\'o}n,
  M.~H. Mu{\~n}oz-Arias, C.~Hellings, A.~Flasby, D.~Colao~Zanuz, J.-C. Besse,
  et~al., Deterministic generation of two-dimensional multi-photon cluster
  states, Nature Communications 16~(1) (2025) 5505.

\bibitem{BCO+09}
S.~Blanes, F.~Casas, J.~A. Oteo, J.~Ros, The magnus expansion and some of its
  applications, Physics Reports 470 (2009) 151.

\bibitem{KLM+16}
V.~Khemani, A.~Lazarides, R.~Moessner, S.~L. Sondhi, Phase structure of driven
  quantum systems, Phys. Rev. Lett. 116 (2016) 250401.

\bibitem{FWV+22}
A.~J. Friedman, B.~Ware, R.~Vasseur, A.~C. Potter, Topological edge modes
  without symmetry in quasiperiodically driven spin chains, Phys. Rev. B 105
  (2022) 115117.

\bibitem{PPS+17}
I.-D. Potirniche, A.~C. Potter, M.~Schleier-Smith, A.~Vishwanath, N.~Y. Yao,
  Floquet symmetry-protected topological phases in cold-atom systems, Phys.
  Rev. Lett. 119 (2017) 123601.

\bibitem{KDP18}
A.~Kumar, P.~T. Dumitrescu, A.~C. Potter, String order parameters for
  one-dimensional floquet symmetry protected topological phases, Phys. Rev. B
  97 (2018) 224302.

\bibitem{MCD+21}
C.~Monroe, W.~C. Campbell, L.-M. Duan, Z.-X. Gong, A.~V. Gorshkov, P.~W. Hess,
  R.~Islam, K.~Kim, N.~M. Linke, G.~Pagano, P.~Richerme, C.~Senko, N.~Y. Yao,
  Programmable quantum simulations of spin systems with trapped ions, Rev. Mod.
  Phys. 93 (2021) 025001.

\bibitem{OK19}
T.~Oka, S.~Kitamura, Floquet engineering of quantum materials, Annual Review of
  Condensed Matter Physics 10~(Volume 10, 2019) (2019) 387--408.

\bibitem{Got98}
D.~Gottesman, Theory of fault-tolerant quantum computation, Phys. Rev. A 57
  (1998) 127--137.

\bibitem{BE21}
N.~P. Breuckmann, J.~N. Eberhardt, Quantum low-density parity-check codes, PRX
  Quantum 2 (2021) 040101.

\bibitem{CM18}
K.~R. Colladay, E.~J. Mueller, Rewiring stabilizer codes, New J. Phys. 20
  (2018) 083030.

\bibitem{BEG+24}
D.~Bluvstein, S.~J. Evered, A.~A. Geim, et~al., Logical quantum processor based
  on reconfigurable atom arrays, Nature 58 (2024) 626.

\bibitem{ABB+24}
C.~Ryan-Anderson, N.~C. Brown, C.~H. Baldwin, et~al., High-fidelity and
  fault-tolerant teleportation of a logical qubit using transversal gates and
  lattice surgery on a trapped-ion quantum computer, Science 385 (2024) 1327.

\bibitem{LB13}
D.~Lidar, T.~A. Brun (Eds.), Quantum error correction, Cambridge University
  Press, 2013.

\bibitem{FP14}
A.~J. Ferris, D.~Poulin, Tensor networks and quantum error correction, Phys.
  Rev. Lett. 113 (2014) 030501.

\bibitem{LW25}
Y.-D. Liu, D.-S. Wang, General channel capacities from channel-state duality,
  arXiv preprint arXiv:2504.01411 (2025).

\bibitem{ZCZ+15}
B.~Zeng, X.~Chen, D.-L. Zhou, X.-G. Wen, Quantum Information Meets Quantum
  Matter, Springer-Verlag New York, 2019.

\bibitem{Kit01}
A.~Y. Kitaev, Unpaired majorana fermions in quantum wires, Phys.-Usp. 44 (2001)
  31.

\bibitem{ZLW19}
H.~Zhang, D.~Liu, M.~Wimmer, et~al., Next steps of quantum transport in
  majorana nanowire devices, Nat. Commun. 10 (2019) 5128.

\bibitem{Mic25}
{Microsoft Quantum}, Roadmap to fault tolerant quantum computation using
  topological qubit arrays, arXiv preprint arXiv:2502.12252 (2025).

\bibitem{ZDJ16}
H.~Zheng, A.~Dua, L.~Jiang, Measurement-only topological quantum computation
  without forced measurements, New J. Phys. 18~(12) (2017) 123027.

\bibitem{NSS+08}
C.~Nayak, S.~H. Simon, A.~Stern, M.~Freedman, S.~D. Sarma, Non-abelian anyons
  and topological quantum computation, Rev. Mod. Phys. 80~(3) (2008) 1083.

\bibitem{SCP10}
N.~Schuch, J.~I. Cirac, D.~Perez-Garcia, {PEPS as ground states: Degeneracy and
  topoloy}, Ann. Phys. 325 (2010) 2153.

\bibitem{BBB21}
S.~Bartolucci, P.~Birchall, H.~Bombin, H.~Cable, C.~Dawson, M.~Gimeno-Segovia,
  E.~Johnston, K.~Kieling, N.~Nickerson, M.~Pant, F.~Pastawski, T.~Rudolph,
  C.~Sparrow, {Fusion-based quantum computation}, Nat. Commun. 14 (2023) 912.

\bibitem{MW23}
R.~Ma, C.~Wang, Average symmetry-protected topological phases, Phys. Rev. X 13
  (2023) 031016.

\bibitem{DZLY25}
K.~Ding, H.-R. Zhang, B.-T. Liu, S.~Yang, Boundary anomaly detection in
  two-dimensional subsystem symmetry-protected topological phases, Phys. Rev. B
  111 (2025) 205125.

\end{thebibliography}
\bibliographystyle{elsarticle-num}

\end{document}